\documentclass[journal]{IEEEtran}
\pdfoutput=1 

\usepackage{etoolbox}
\usepackage{csquotes} 
\usepackage[shortcuts]{glossaries} 
\glsenableentrycount 
\usepackage[dvipsnames]{xcolor} 
\usepackage{amsthm} 

\usepackage{orcidlink}

\usepackage{tikz}
\usetikzlibrary{
    positioning,
    calc,
    math,
    plotmarks,
    fit,
    matrix,
}
\usepackage{pgfplots}
\pgfplotsset{
    compat=newest, 
    every axis/.style={
        /pgf/number format/set thousands separator={}, 
    },
}

\usepackage{siunitx}
\DeclareSIUnit\dBm{dBm} 

\usepackage[capitalize]{cleveref} 

\usepackage{currfile}
\ProvideDocumentCommand{\CurrentFilePath}{}{\currfiledir}

\NewDocumentCommand{\crefSubfig}{mm}{\cref{#1}(#2)}
\NewDocumentCommand{\CrefSubfig}{mm}{\Cref{#1}(#2)}

\definecolor{colA}{RGB}{ 31,  120,  180}
\definecolor{colB}{RGB}{231,  41,   138}
\definecolor{colC}{RGB}{ 27,  158,  119}
\definecolor{colD}{RGB}{217,  95,     2}
\definecolor{colE}{RGB}{118,  42,   131}
\definecolor{colF}{RGB}{230,  171,    2}

\NewDocumentCommand{\mA}{}{{\ensuremath{\textcolor{colA}{\mathrm{A}}}}}
\NewDocumentCommand{\mB}{}{{\ensuremath{\textcolor{colB}{\mathrm{B}}}}}
\NewDocumentCommand{\mC}{}{{\ensuremath{\textcolor{colC}{\mathrm{C}}}}}
\NewDocumentCommand{\mD}{}{{\ensuremath{\textcolor{colD}{\mathrm{D}}}}}
\NewDocumentCommand{\mX}{}{{\ensuremath{\textcolor{colE}{\mathrm{X}}}}}

\NewDocumentCommand{\qfwmBw }{}{\ensuremath{{ B_\mathrm{FWM}      }}} 
\NewDocumentCommand{\qfwmEff}{}{\ensuremath{{ \eta_{\mathrm{FWM}} }}} 

\newlength\figW

\NewDocumentCommand{\TE}{m}{{\ensuremath{\mathrm{TE}_{#1}}}}
\NewDocumentCommand{\TM}{m}{{\ensuremath{\mathrm{TM}_{#1}}}}
\NewDocumentCommand{\HE}{m}{{\ensuremath{\mathrm{HE}_{#1}}}}
\NewDocumentCommand{\EH}{m}{{\ensuremath{\mathrm{EH}_{#1}}}}

\NewDocumentCommand{\TODO}{d()}{
    \textcolor{red}{
        \textbf{TODO}
        \IfValueT{#1}{~(#1)}
    }
}
\NewDocumentCommand{\rev}{+m}{%
    #1%
}

\pgfplotsset{colormap={parula}{
    rgb255=(53,42,135)
    rgb255=(15,92,221)
    rgb255=(18,125,216)
    rgb255=(7,156,207)
    rgb255=(21,177,180)
    rgb255=(89,189,140)
    rgb255=(165,190,107)
    rgb255=(225,185,82)
    rgb255=(252,206,46)
    rgb255=(249,251,14)
}}

\theoremstyle{definition}
\newtheorem{definition}{Definition}

\newrobustcmd{\conj}{{*}}

\DeclareMathOperator{\nn}{n}
\DeclareMathOperator{\cc}{c}
\DeclareMathOperator{\LL}{L}
\DeclareMathOperator{\PP}{P}

\renewcommand{\Vec}[1]{\vec{#1}}

\newrobustcmd{\derivative}[4]{%
    \ifstrempty{#2}%
        {\left. \frac{#1#3}{#1#4}             \right.}%
        {\left. \frac{#1^{#2}\,#3}{#1#4^{#2}} \right.}%
}

\newrobustcmd{\dd}[3][]{\derivative{\mathrm{d}}{#1}{#2}{#3}}

\newrobustcmd{\real}[1]{\mathcal{R} e \{ #1 \}}

\newrobustcmd{\abs}[1]{\left\vert #1 \right\vert}

\newcommand{\expp}[1]{\operatorname{e}^{#1}}
\newcommand{\exps}[1]{\exp\left(#1\right)}

\newacronym{aowlc}{AOWC}{all-optical wavelength converter}
\newacronym{ber}{BER}{bit error rate}
\newacronym{bs}{BS}{Bragg scattering}
\newacronym{b2b}{B2B}{back to back}
\newacronym{cd}{CD}{chromatic dispersion}
\newacronym{ce}{CE}{input-output conversion efficiency}
\newacronym{cspr}{CSPR}{carrier-to-signal power ratio}
\newacronym{cw}{CW}{continuous wave}
\newacronym{dsp}{DSP}{digital signal processing}
\newacronym{ecl}{ECL}{external-cavity laser}
\newacronym{edfa}{EDFA}{Erbium-doped fiber amplifier}
\newacronym{epic}{EPIC}{electronic and photonic integrated circuit}
\newacronym{fca}{FCA}{free-carrier absorption}
\newacronym{fdm}{FDM}{finite difference method}
\newacronym{fmf}{FMF}{few-mode fiber}
\newacronym{fwm}{FWM}{four-wave mixing}
\newacronym{1fwm}{1-FWM}{one-mode four-wave mixing}
\newacronym{2fwm}{2-FWM}{two-mode four-wave mixing}
\newacronym{12fwm}{1/2-FWM}{one- or two-mode four-wave mixing}
\newacronym{3fwm}{3-FWM}{three-mode four-wave mixing}
\newacronym{4fwm}{4-FWM}{four-mode four-wave mixing}
\newacronym{34fwm}{3/4-FWM}{three- or four-mode four-wave mixing}
\newacronym{gc}{GC}{grating coupler}
\newacronym{hdfec}{HDFEC}{hard-decision forward error correction}
\newacronym{hnlf}{HNLF}{highly nonlinear fiber}
\newacronym{il}{IL}{insertion loss}
\newacronym{kkrx}{KK-Rx}{Kramers-Kronig receiver}
\newacronym{madm}{MADM}{mode add-drop multiplexer}
\newacronym{mmwg}{MMWG}{multi-mode waveguide}
\newacronym{nr}{NR}{nano-rib}
\newacronym{opc}{OPC}{optical phase conjugation}
\newacronym{osnr}{OSNR}{optical signal to noise ratio}
\newacronym{osa}{OSA}{optical spectrum analyzer}
\newacronym{pd}{PD}{photo diode}
\newacronym{pdfa}{PDFA}{Praseodymium-doped fiber amplifier}
\newacronym{pic}{PIC}{photonic integrated circuit}
\newacronym{pdla}{PDLA}{photonic dispersion and loss analyzer}
\newacronym{pm}{PM}{phase matching}
\newacronym{qpsk}{QPSK}{quadrature phase-shift keying}
\newacronym{roadm}{ROADM}{reconfigurable optical add-drop multiplexer}
\newacronym{sdfec}{SDFEC}{soft-decision forward error correction}
\newacronym{sdm}{SDM}{space-division multiplex}
\newacronym{smwg}{SMWG}{single-mode waveguide}
\newacronym{soa}{SOA}{semiconductor optical amplifier}
\newacronym{soi}{SOI}{silicon on insulator}
\newacronym{ssmf}{SSMF}{standard single-mode fiber}
\newacronym{tobpf}{T-OBPF}{tunable optical band-pass filter}
\newacronym{tpa}{TPA}{two-photon absorption}
\newacronym{ubpsp}{UB-PSP}{ultra-broadband photonic signal processor}
\newacronym{voa}{VOA}{variable optical attenuator}
\newacronym{wlc}{WLC}{wavelength conversion}
\newacronym{wdm}{WDM}{wavelength-division multiplexing}
\newacronym{wss}{WSS}{wavelength-selective switch}
\newacronym{xt}{XT}{crosstalk}

\begin{document}
\NewDocumentCommand{\myTitle}{}{\rev{Phase Matching for} Multimode Four-Wave Mixing in Few-Mode Fibers and Nano-Rib Waveguides}

\title{\myTitle{}}

\author{%
    Tasnad~Kernetzky\,\orcidlink{0000-0002-5111-5938} and~Norbert~Hanik%
    \thanks{The authors are with Technical University of Munich.}%
    \thanks{E-mails: \mbox{tasnad@tum.de} and norbert.hanik@tum.de.}%
    \thanks{This work was supported by DFG project HA~6010/6-1.}%
}

\maketitle

\begin{abstract}
We compare \rev{\acrlong{pm} for} \acrlong{fwm} using one, two, three, and four waveguide modes.
For the comparison, we use numerical optimizations \rev{and an estimate of the generated idler power}.
We present results for \acrlongpl{fmf} and \acrlong{nr} waveguides and show that for both waveguide types, \acrlong{fwm} bandwidths and idler powers are best for one- and two-mode operation and that \acrlong{4fwm} is not feasible at all.
Some \acrlong{nr} waveguides support three-mode \acrlong{fwm}, albeit with much reduced bandwidth and reduced idler power.
\end{abstract}

\begin{IEEEkeywords}
nonlinear optics, phase matching, multimode waveguide, four-wave mixing, few-mode fiber, SOI nano-rib waveguide
\end{IEEEkeywords}

\section{Introduction}
\cGls{fwm} is usually regarded as a detrimental effect for optical communications over long fiber links.
Although silica has a relatively low nonlinearity coefficient $\gamma$, the long propagation distance of up to thousands of kilometers leads to non-negligible nonlinear distortions.
There exist different approaches for mitigating these effects, one of which is \cgls{opc}.
There, \cgls{fwm} is intentionally used in a lumped device in the middle of the optical link to cancel linear and nonlinear phase distortions caused by \cgls{cd} and fiber nonlinearity.
A second use case for \cgls{fwm} utilizes the so-called \cgls{bs} process to all-optically shift the frequency of transmission channels or even bands.

To process an optical signal by means of \cgls{fwm}, it is sent into a cubic nonlinear medium, together with two strong pumps.
If laser frequencies, waveguide geometry and choice of modes are well-selected, the nonlinearity will generate an idler wave with the desired properties (conjugated phase and/or shifted frequency).
Typically, two configurations are used for waveguide modes and laser frequencies: \cgls{1fwm} and \cgls{2fwm}.
In \cgls{1fwm}, all lasers (signal and pumps) are launched into the same mode, and the idler will be generated in the same mode as well.
For \cgls{pm}, frequencies need to be close to a zero dispersion region of the waveguide (e.g., \cite{hillCWThreewaveMixing1978}).
In \cgls{2fwm}, the signal and one pump are launched into one mode, while the other pump is launched into another mode where also the idler will be generated (e.g., \cite{essiambreExperimentalInvestigationInterModal2013}).
\rev{In \cgls{3fwm}, signal, pumps and idler propagate in exactly three different modes (two modes with one wave, one mode with two waves).
Finally, in \cgls{4fwm}, all waves propagate in four different modes.
While there exist analytical expressions for \cgls{pm} in \cgls{1fwm} and \cgls{2fwm}, numerical optimizations are necessary for \cgls{3fwm} and \cgls{4fwm}.}

\rev{In this work, we compare \cgls{fwm} bandwidths (by a metric $\qfwmBw$, see \cref{thm:fwmBw}) and nonlinearity parameters $\gamma$, for the cases of \cgls{1fwm}, \cgls{2fwm}, \cgls{3fwm} and \cgls{4fwm}.}
We consider graded index depressed cladding \cglspl{fmf} and \cgls{nr} silicon waveguides, both with different geometries.
\rev{We chose these two types of waveguides, since they are widely used for all-optical signal processing (e.g. \cite{essiambreExperimentalInvestigationInterModal2013,koosNonlinearSilicononinsulatorWaveguides2007}).}

\rev{In \cite{ronnigerEfficientUltrabroadbandCtoO2021,kernetzkyNumericalOptimizationCW2021}, we presented a \cgls{2fwm} system experiment which demonstrates all-optical conversion of three channels from C- to O-band.
The extension of all-optical signal processing to more than two modes in future \cgls{sdm} networks could potentially be beneficial.
Hence, we explore its feasibility in this paper.}

\IEEEpubidadjcol 
\section{Four-Wave Mixing and Phase Matching}
\begin{figure}[tb]
    \centering
    \footnotesize
    \setlength\figW{0.7\linewidth} \input{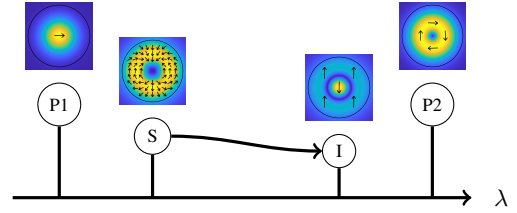}
    \caption{Example of an \cgls{opc} \cgls{4fwm} process. Note that this figure is merely an example and modes are not selected according to any specific condition.}
    \label{fig:fwmFour}
\end{figure}
The principle of multimode \cgls{fwm} is sketched in \cref{fig:fwmFour}.
It shows an example of a four-mode \cgls{opc} operation \rev{(each wave propagates in a different mode)} with pumps P\textsubscript{1} and P\textsubscript{2}, signal S and idler I.

Photon energy ($\hbar\omega$) and photon momentum ($\hbar\beta$) conservation for the \cgls{bs} and \cgls{opc} processes lead to
\begin{align}
    \text{\acrshort{bs}}: \, &\omega_\mathrm{I} = \omega_{\mathrm{P}_1} + \omega_\mathrm{S} - \omega_{\mathrm{P}_2}\label{eq:fwmEnergyConservationBs}\\
    &\Delta\beta      = \beta^\mA(\omega_{\mathrm{P}_1}) + \beta^\mB(\omega_\mathrm{S}) - \beta^\mC(\omega_{\mathrm{P}_2}) - \beta^\mD(\omega_\mathrm{I})\label{eq:fwmDeltaBetaBs}\\
    \text{\acrshort{opc}}: \, &\omega_\mathrm{I} = \omega_{\mathrm{P}_1} - \omega_\mathrm{S} + \omega_{\mathrm{P}_2}\label{eq:fwmEnergyConservationPc}\\
    &\Delta\beta      = \beta^\mA(\omega_{\mathrm{P}_1}) - \beta^\mB(\omega_\mathrm{S}) + \beta^\mC(\omega_{\mathrm{P}_2}) - \beta^\mD(\omega_\mathrm{I})\label{eq:fwmDeltaBetaPc},
\end{align}
where \mA{}, \mB{}, \mC{} and \mD{} denote different modes and, e.g., $\beta^\mA$ is the propagation constant of mode \mA{}.
Firstly, energy conservation dictates the resulting idler frequency as a function of the input laser frequencies (\cref{eq:fwmEnergyConservationBs,eq:fwmEnergyConservationPc}).
Secondly, the efficiency of \cgls{fwm} depends on how well the momentum conservation equation is fulfilled, which is measured by the residual phase mismatch $\Delta\beta$ (\cref{eq:fwmDeltaBetaBs,eq:fwmDeltaBetaPc}).
\rev{For increasing values of phase mismatch, the coherent buildup of the idler turns into an oscillation with decreasing period.}
Minimizing phase mismatch is the well-known process of \textit{\acrlong{pm}} and our approach is explained in \cite{kernetzkyMultiDimensionalOptimization2020} in more detail.
Since modes have different propagation constants and waveguide geometry affects modes, both factors play important roles in \cgls{pm}.

The conclusions we derive in this article \rev{are the same} for \cgls{bs} and \cgls{opc}.
\rev{Therefore, we only show results for \cgls{bs} and mention small differences to \cgls{opc} briefly in the text.}

A useful and intuitive \cgls{pm} approach for \cgls{1fwm} and \cgls{2fwm} was given in \cite{essiambreExperimentalInvestigationInterModal2013}.
The group delay (see \cref{eq:groupDelayAndDispersion}) of signal and pump in one mode at their average frequency, needs to match the group delay in the other mode at the average frequency of idler and the other pump, e.g.,
\begin{align}
    \tau_g^\mA\left(\frac{\omega_{\mathrm{P}_1} + \omega_\mathrm{S}}{2}\right) &= \tau_g^\mA\left(\frac{\omega_{\mathrm{P}_2} + \omega_\mathrm{I}}{2}\right) \quad \text{(\cgls{1fwm})}\label{eq:pmApproxOneMode}\\
    \tau_g^\mA\left(\frac{\omega_{\mathrm{P}_1} + \omega_\mathrm{S}}{2}\right) &= \tau_g^\mB\left(\frac{\omega_{\mathrm{P}_2} + \omega_\mathrm{I}}{2}\right) \quad \text{(\cgls{2fwm})}\label{eq:pmApproxTwoMode}.
\end{align}
This approach is not applicable in more-than-two-mode \cgls{fwm}, as for instance in the example in \cref{fig:fwmFour}.
In the general case, numerical optimizations of \cref{eq:fwmEnergyConservationBs,eq:fwmDeltaBetaBs} need to be performed to find the optima.

Since we are interested in broadband operation, we use the \cgls{fwm} bandwidth as metric of quality.
\begin{definition}\label{thm:fwmBw}
    The \cgls{fwm} bandwidth $\qfwmBw$ is the frequency range the signal can be moved, while the \rev{estimate of} idler power does not drop by more than \qty{3}{\dB} from its peak value, and without changing any other parameter (waveguide dimensions, pump frequencies, mode assignments).
\end{definition}

The propagation constant in a mode \mA{} can be expanded into its Taylor series
\begin{align}
    \beta^{\mA}(\omega) &= \beta^{\mA}_0 + \beta^{\mA}_1\Delta\omega + \frac{1}{2}\beta^{\mA}_2 \Delta\omega^2 + \dots,\label{eq:betaTaylor}\\
          \beta^{\mA}_n &= \left.\dd[n]{\beta^{\mA}(\omega)}{\omega}\right\vert_{\omega = \omega_0}, \quad \Delta\omega = \omega - \omega_0
\end{align}
around a center frequency $\omega_0$.
The $\beta_0$ coefficients are usually subject to fluctuations along fiber waveguides.
Therefore, it is customary in \cgls{2fwm} to choose the modes such that the $\beta_0$s are canceled in \cref{eq:fwmDeltaBetaBs} (see, e.g., \cite{essiambreExperimentalInvestigationInterModal2013}) in order to ensure disturbance-free \cgls{fwm}.
However, it requires further studies to determine if this assumption also holds for short \cgls{nr} waveguides with propagation distances in the range of centimeters.
In \cgls{1fwm}, the $\beta_0$s cancel automatically and in \cgls{2fwm} they can be canceled by proper mode selection. However, there is no way to select modes in \cgls{3fwm} and \cgls{4fwm} to cancel them.
Thus, we ignore this limitation in the current work and allow all mode combinations.
This way, we ensure a fair comparison between \cgls{fwm} with one to four modes.

\section{Geometry, Modes and Dispersion \rev{Properties} of Few-Mode Fibers and Nano-Rib Waveguides}\label{sec:fmfNr}
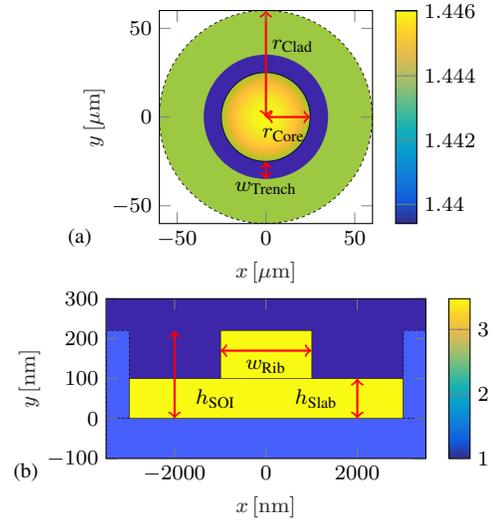
\begin{figure}[tb]
    \centering
    \footnotesize
    \setlength\figW{0.8\linewidth} \begin{tikzpicture}
\begin{axis}[
    name = nFmfPic,
    enlargelimits=false,
    axis on top,
    width=0.4\figW,
    height=0.4\figW,
    scale only axis,
    xlabel style={font=\color{white!15!black}},
    xlabel={$x \, \mathrm{\left[\mu{}m\right]}$},
    ylabel={$y \, \mathrm{\left[\mu{}m\right]}$},
    ylabel style={font=\color{white!15!black}, yshift=-1mm},
    point meta min=1.4394, point meta max=1.44603,
    colorbar,
    colorbar style={width=8, /pgf/number format/precision=4},
]
\addplot graphics [xmin=-60,xmax=60,ymin=-60,ymax=60] {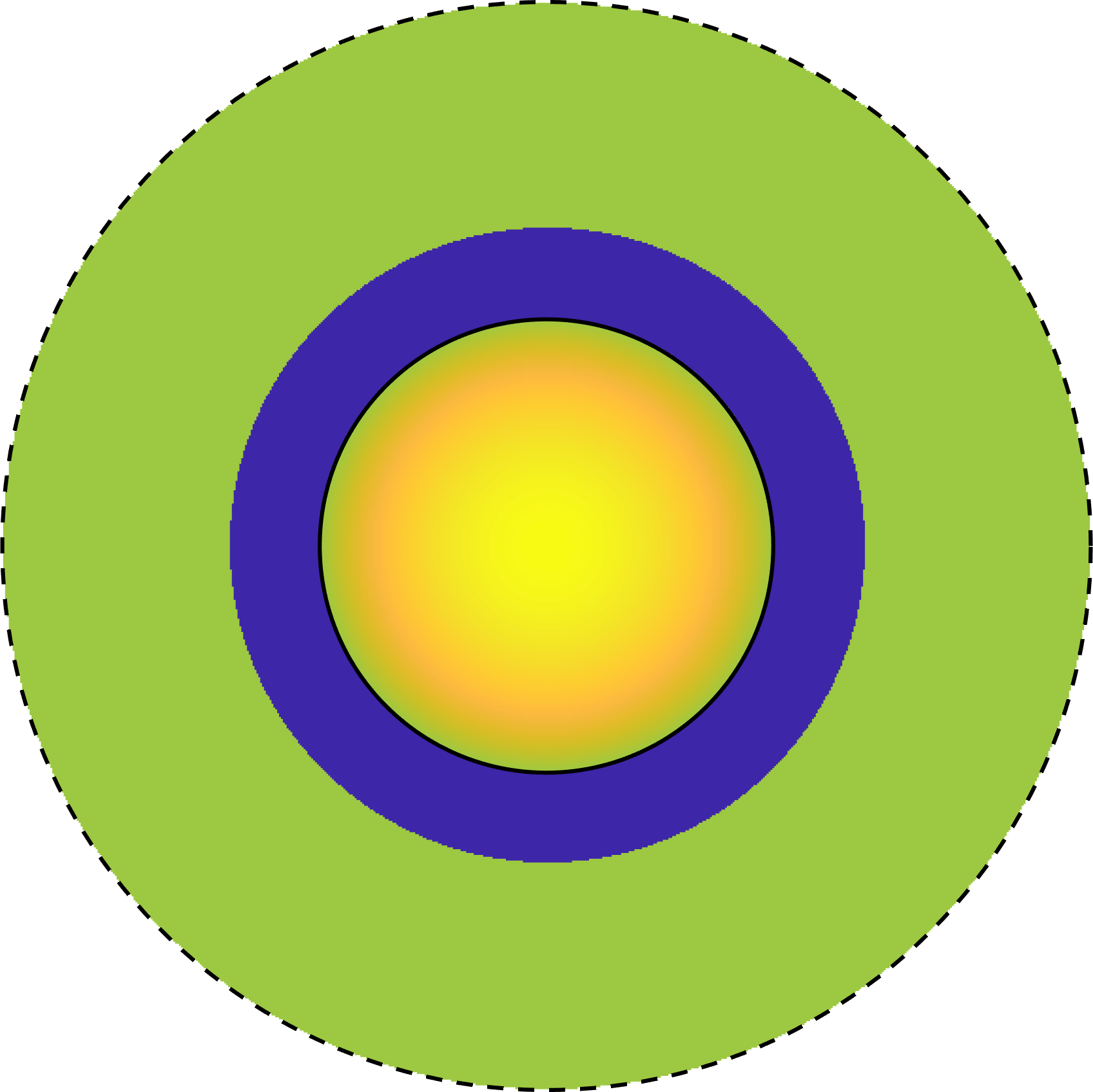};
\draw[red, thick, <->] (axis cs:0,0)  -- (axis cs:0,60);
\node at(axis cs:15,40) {$r_\text{Clad}$};
\draw[red, thick, <->] (axis cs:0,0)  -- (axis cs:25,0);
\node at(axis cs:10,-10) {$r_\text{Core}$};
\draw[red, thick, <->] (axis cs:0,-25)  -- (axis cs:0,-35);
\node at(axis cs:0,-40) {$w_\text{Trench}$};
\end{axis}
\begin{axis}[
    name = nNrPic,
    at = (nFmfPic.south),
    anchor = north,
    yshift = -1cm,
    enlargelimits=false,
    axis on top,
    width=0.6\figW,
    height=0.3\figW,
    scale only axis,
    xlabel style={font=\color{white!15!black}},
    xlabel={$x \, \mathrm{\left[nm\right]}$},
    ylabel={$y \, \mathrm{\left[nm\right]}$},
    ylabel style={font=\color{white!15!black}, yshift=-1mm},
    point meta min=1, point meta max=3.47641,
    colorbar style={width=8},
    colorbar,
]
\addplot graphics [xmin=-3500,xmax=3500,ymin=-100,ymax=300] {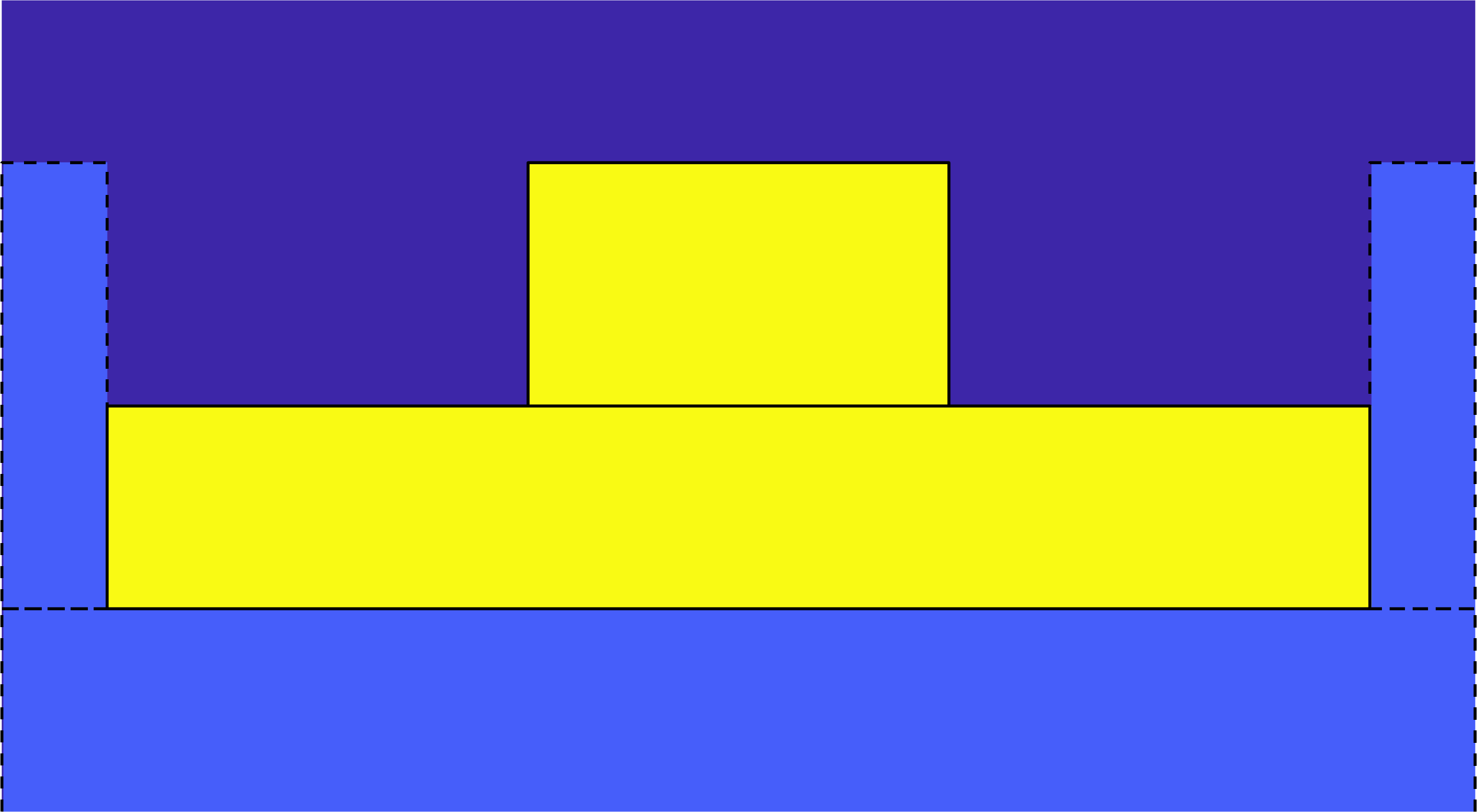};
\node at(axis cs:-1100,50) {$h_\text{SOI}$};
\draw[red, thick, <->] (axis cs:-2000,0)  -- (axis cs:-2000,220);
\node at(axis cs:1100,50) {$h_\text{Slab}$};
\draw[red, thick, <->] (axis cs:2000,0)   -- (axis cs:2000,100);
\node at(axis cs:0,130) {$w_\text{Rib}$};
\draw[red, thick, <->] (axis cs:-1000,170) -- (axis cs:1000,170);
\end{axis}
\node (a label) [left=of nFmfPic.south west, anchor=south east, shift=({0.2, -0.4})] {(a)};
\node (b label)  [left=of nNrPic.south  west, anchor=south east, shift=({0.2, -0.4})] {(b)};
\end{tikzpicture}
    \caption{Refractive index geometries. (a) \cGlspl{fmf} with core radius, cladding radius and trench width. (b) \cGls{nr} waveguides with slab height, \acrshort{soi} height and rib width.}
    \label{fig:n_fmf_nr}
\end{figure}
\begin{table}[tb]
    \centering
    \caption{Geometry values considered for optimizations.}
    \label{tab:simParams}
    \begin{tabular}{ccc}
        \textbf{Parameter}        & \textbf{Values}                                      & \textbf{Number of Values}\\
        \hline
        \multicolumn{3}{l}{\textit{Few-Mode Fiber}}\\
        $r_\text{Core}$           & \qtyrange{6}{40}{\um}                                  & $18$\\
        $w_\text{Trench}$         & $\left\{ 0, 0.25, 0.5, 1, 2, 4, 8 \right\} \unit{\um}$ & $7$\\
        $r_\text{Clad}$           & $r_\text{Core} + w_\text{Trench} + \qty{10}{\um}$      & -\\
        $\rev{\Delta}$            & \rev{\qty{0.1383}{\percent}}                           & 1\\
        \hline
        \multicolumn{3}{l}{\textit{Nano-Rib Waveguide}}\\
        $w_\text{Rib}$            & \qtyrange{1000}{3000}{\nm}                            & $20$\\
        $h_\text{Slab}$           & \qtyrange{70}{180}{\nm}                               & $12$\\
        $h_\text{\acrshort{soi}}$ & $\qty{220}{\nm}$                                      & $1$
    \end{tabular}
\end{table}
\begin{figure}[tb]
    \centering
    \footnotesize
    \setlength\figW{0.97\linewidth} \begin{tikzpicture}
\begin{axis}[
    name=plot_1_1,
    width=0.24\figW,
    height=0.24\figW,
    scale only axis,
    axis on top,
    ticks=none,
    enlargelimits=false,
]
\addplot graphics [xmin=-24, xmax=24, ymin=-24, ymax=24] {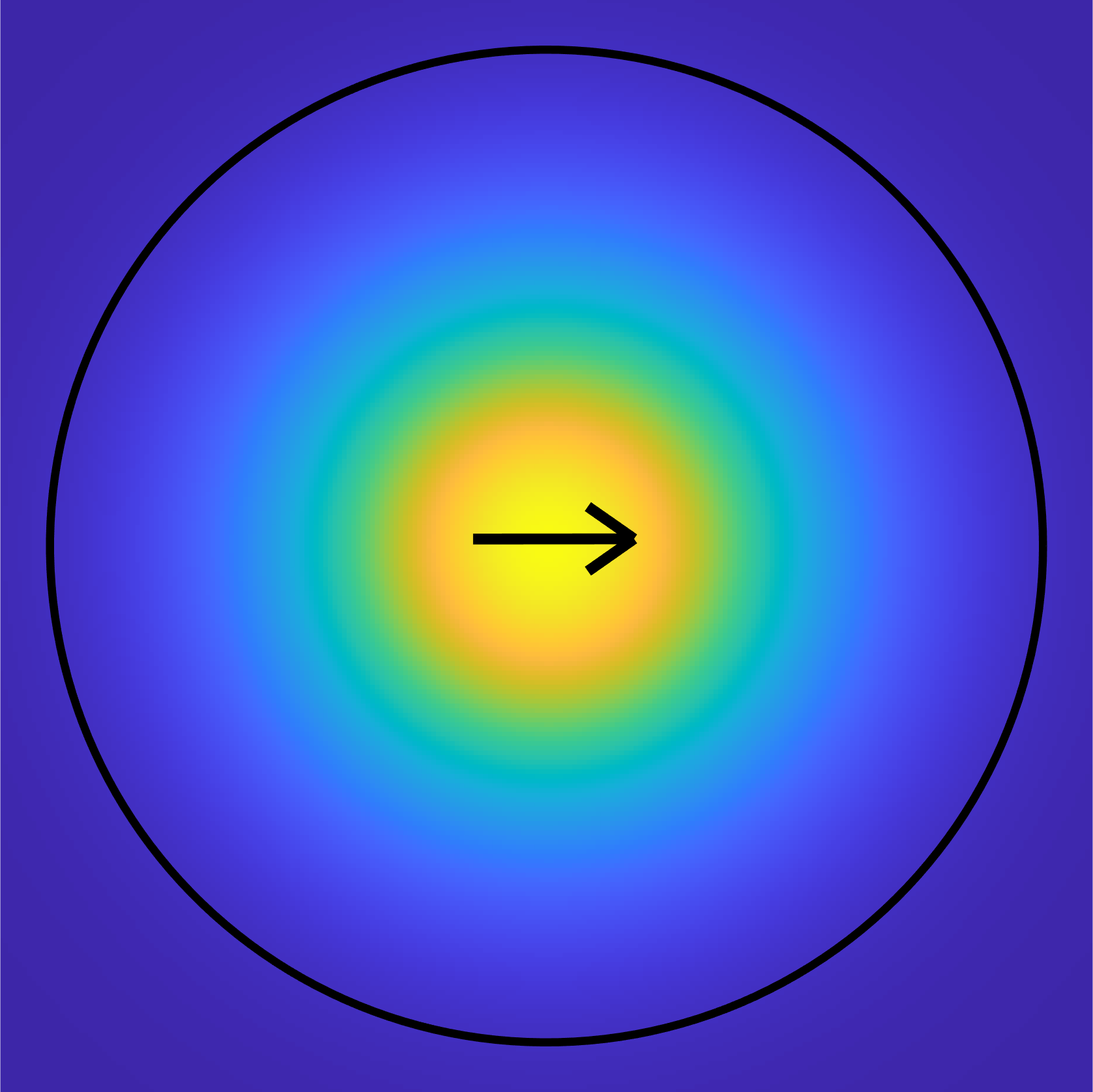};
\end{axis}
\begin{axis}[
    name=plot_1_2,
    at = (plot_1_1.east),
    anchor = west,
    xshift = 0.2cm,
    width=0.24\figW,
    height=0.24\figW,
    scale only axis,
    axis on top,
    ticks=none,
    enlargelimits=false,
]
\addplot graphics [xmin=-24, xmax=24, ymin=-24, ymax=24] {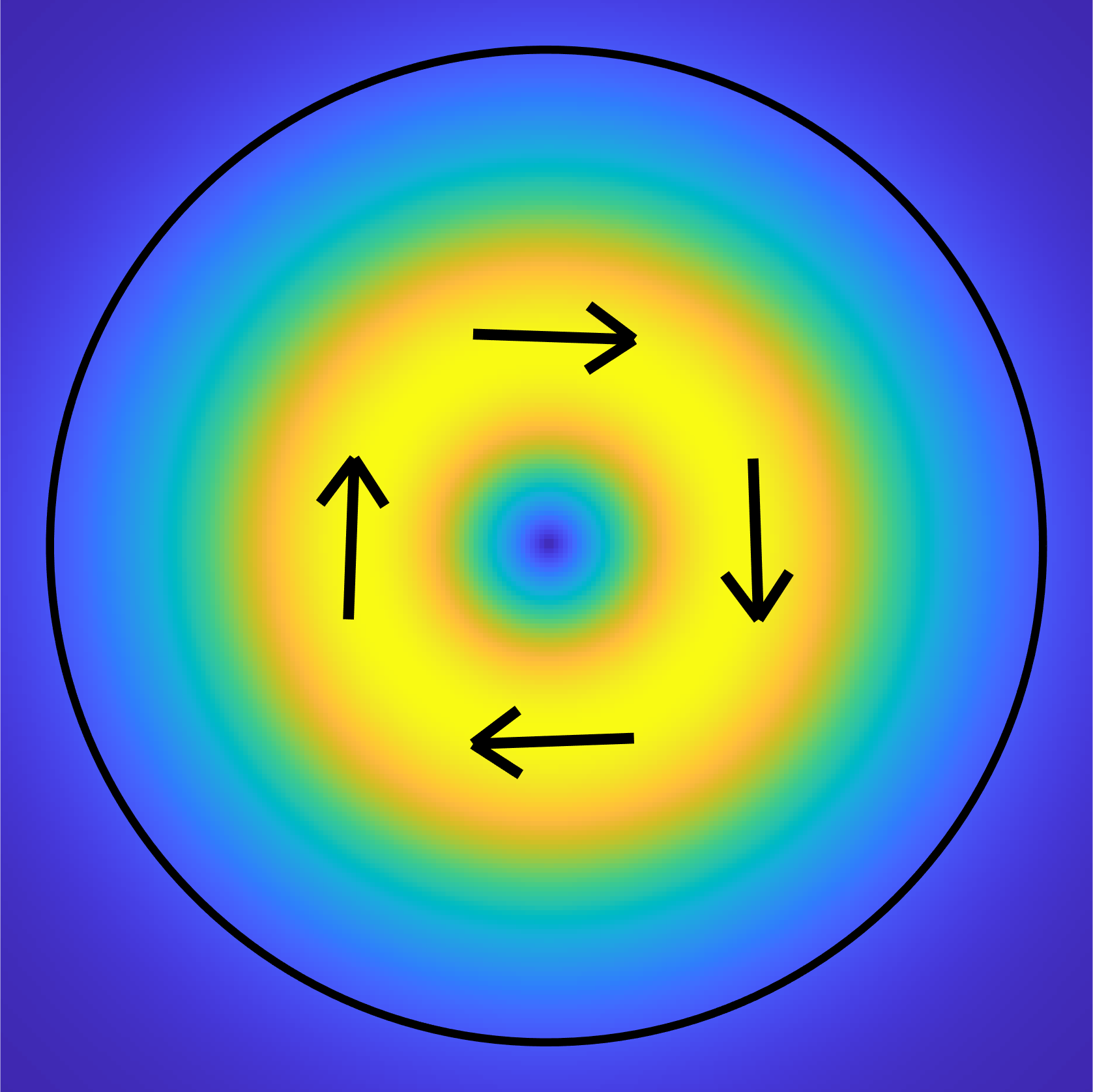};
\end{axis}
\begin{axis}[
    name=plot_1_3,
    at = (plot_1_2.east),
    anchor = west,
    xshift = 0.2cm,
    width=0.24\figW,
    height=0.24\figW,
    scale only axis,
    axis on top,
    ticks=none,
    enlargelimits=false,
]
\addplot graphics [xmin=-24, xmax=24, ymin=-24, ymax=24] {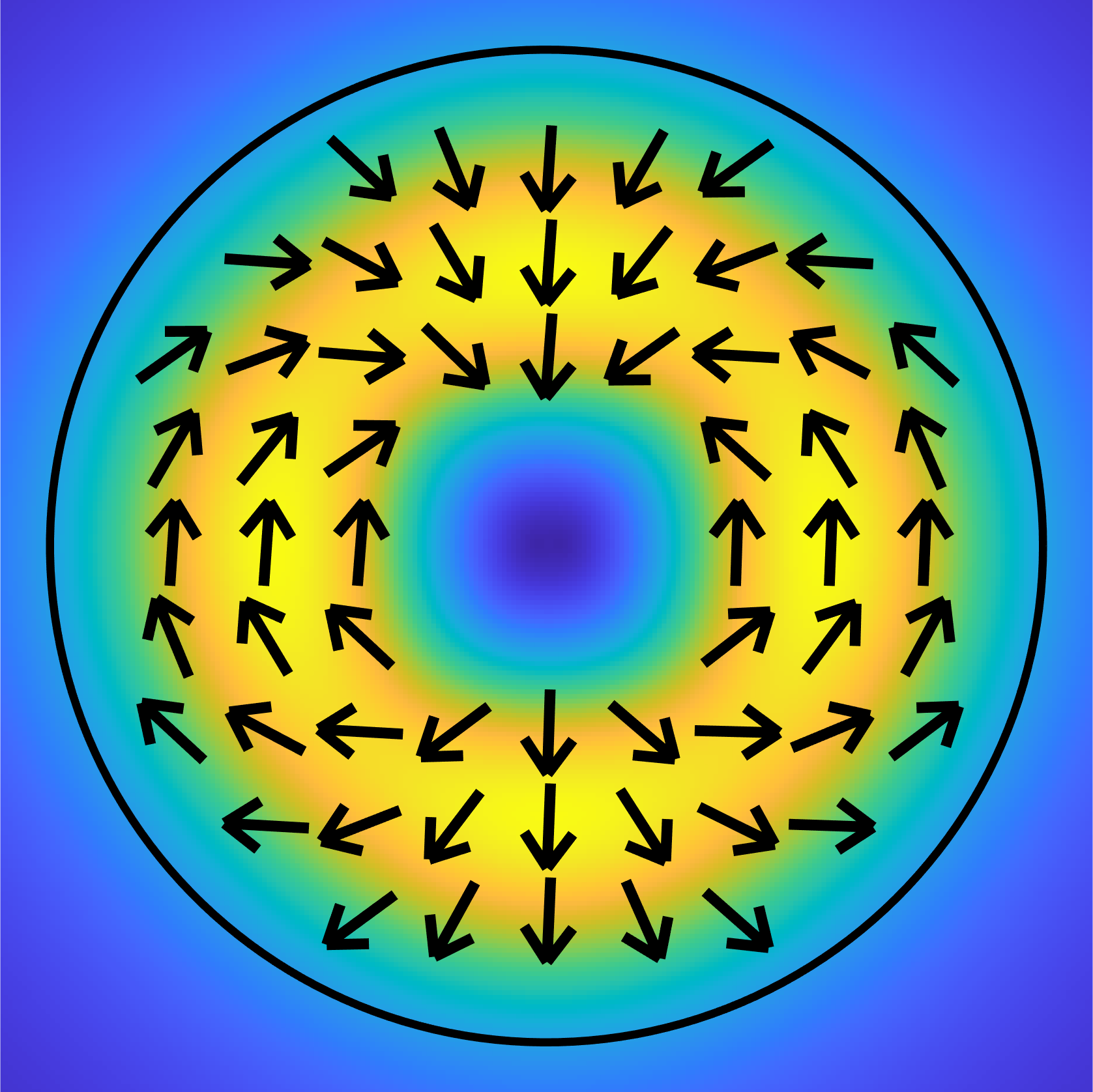};
\end{axis}
\begin{axis}[
    name=plot_1_4,
    at = (plot_1_3.east),
    anchor = west,
    xshift = 0.2cm,
    width=0.24\figW,
    height=0.24\figW,
    scale only axis,
    axis on top,
    ticks=none,
    enlargelimits=false,
]
\addplot graphics [xmin=-24, xmax=24, ymin=-24, ymax=24] {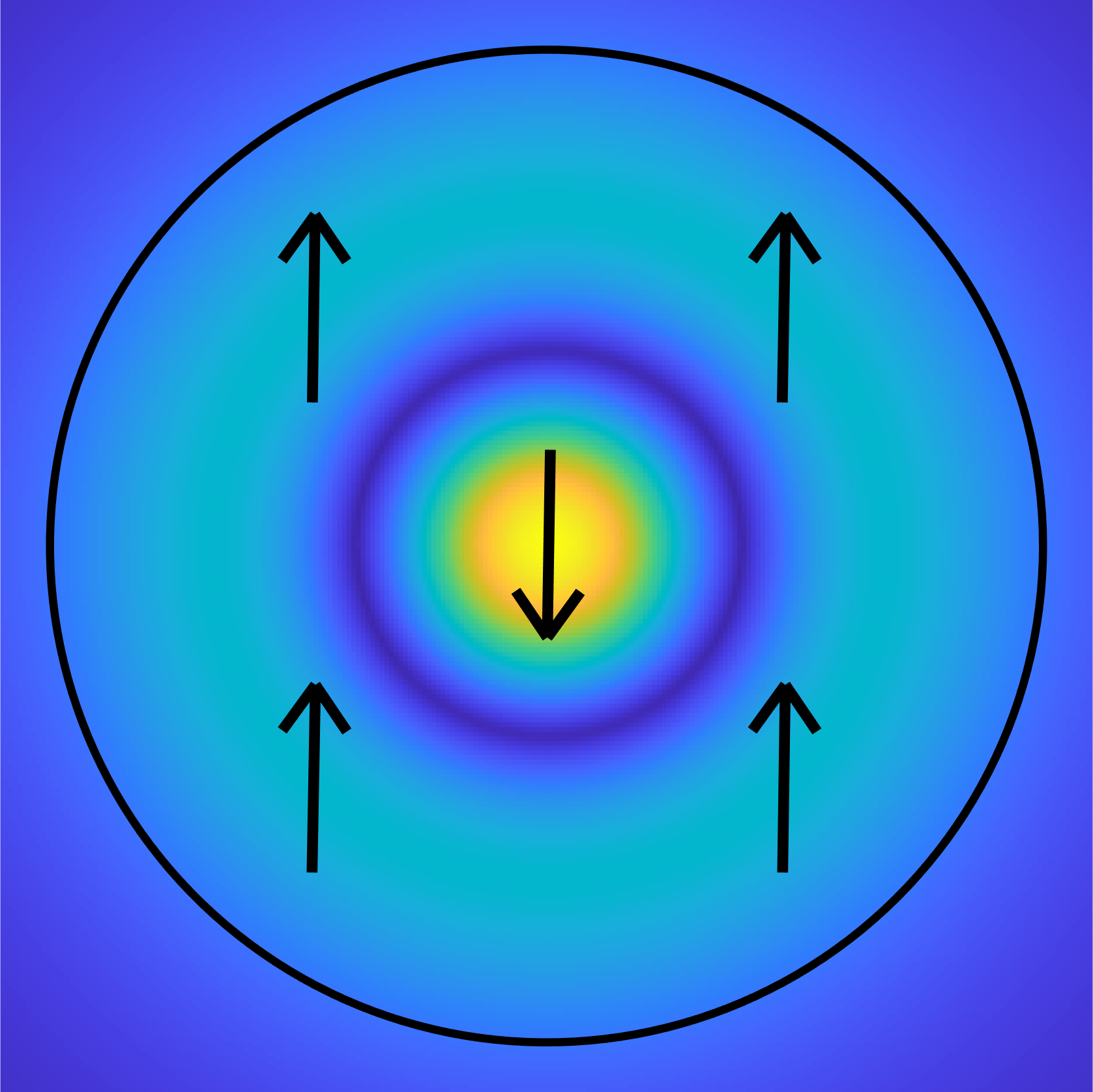};
\end{axis}
\begin{axis}[
    name=plot_2_1,
    at = (plot_1_1.south),
    anchor = north,
    yshift = -0.2cm,
    width=0.24\figW,
    height=0.24\figW*0.8,
    scale only axis,
    axis on top,
    ticks=none,
    enlargelimits=false,
]
\addplot graphics [xmin=-1168.422,xmax=1168.422,ymin=-200,ymax=420] {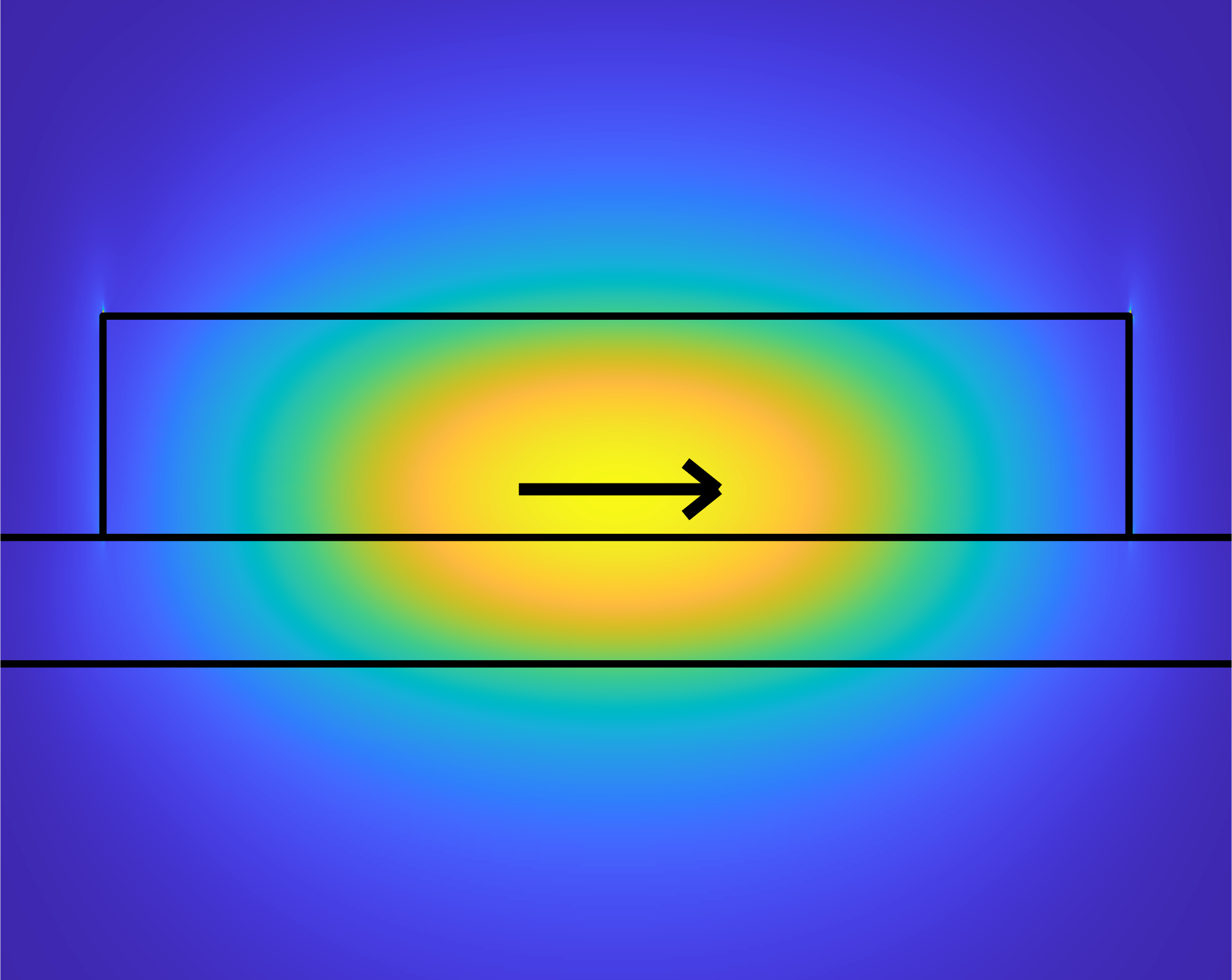};
\end{axis}
\begin{axis}[
    name=plot_2_2,
    at = (plot_2_1.east),
    anchor = west,
    xshift = 0.2cm,
    width=0.24\figW,
    height=0.24\figW*0.8,
    scale only axis,
    axis on top,
    ticks=none,
    enlargelimits=false,
]
\addplot graphics [xmin=-1168.422,xmax=1168.422,ymin=-200,ymax=420] {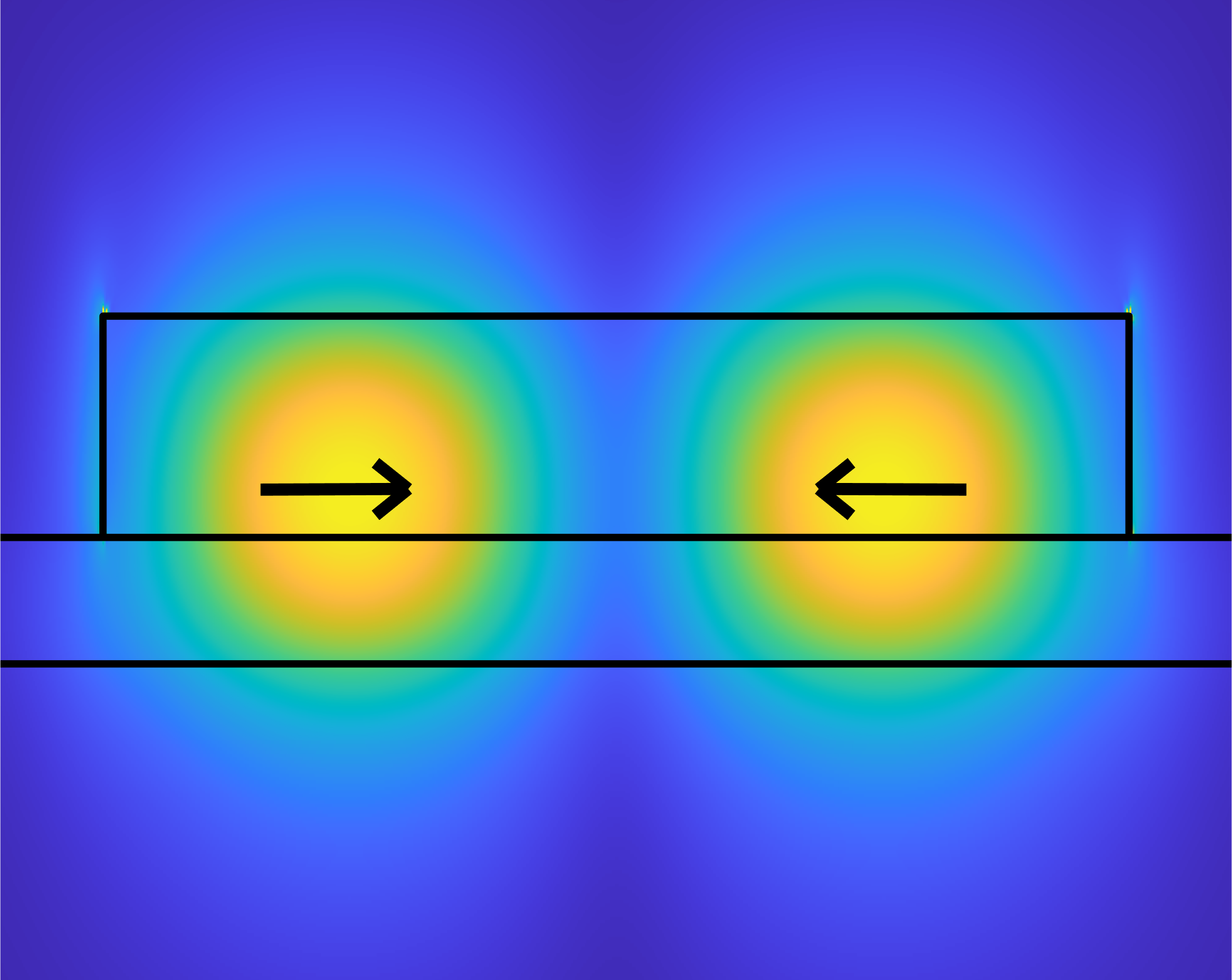};
\end{axis}
\begin{axis}[
    name=plot_2_3,
    at = (plot_2_2.east),
    anchor = west,
    xshift = 0.2cm,
    width=0.24\figW,
    height=0.24\figW*0.8,
    scale only axis,
    axis on top,
    ticks=none,
    enlargelimits=false,
]
\addplot graphics [xmin=-1168.422,xmax=1168.422,ymin=-200,ymax=420] {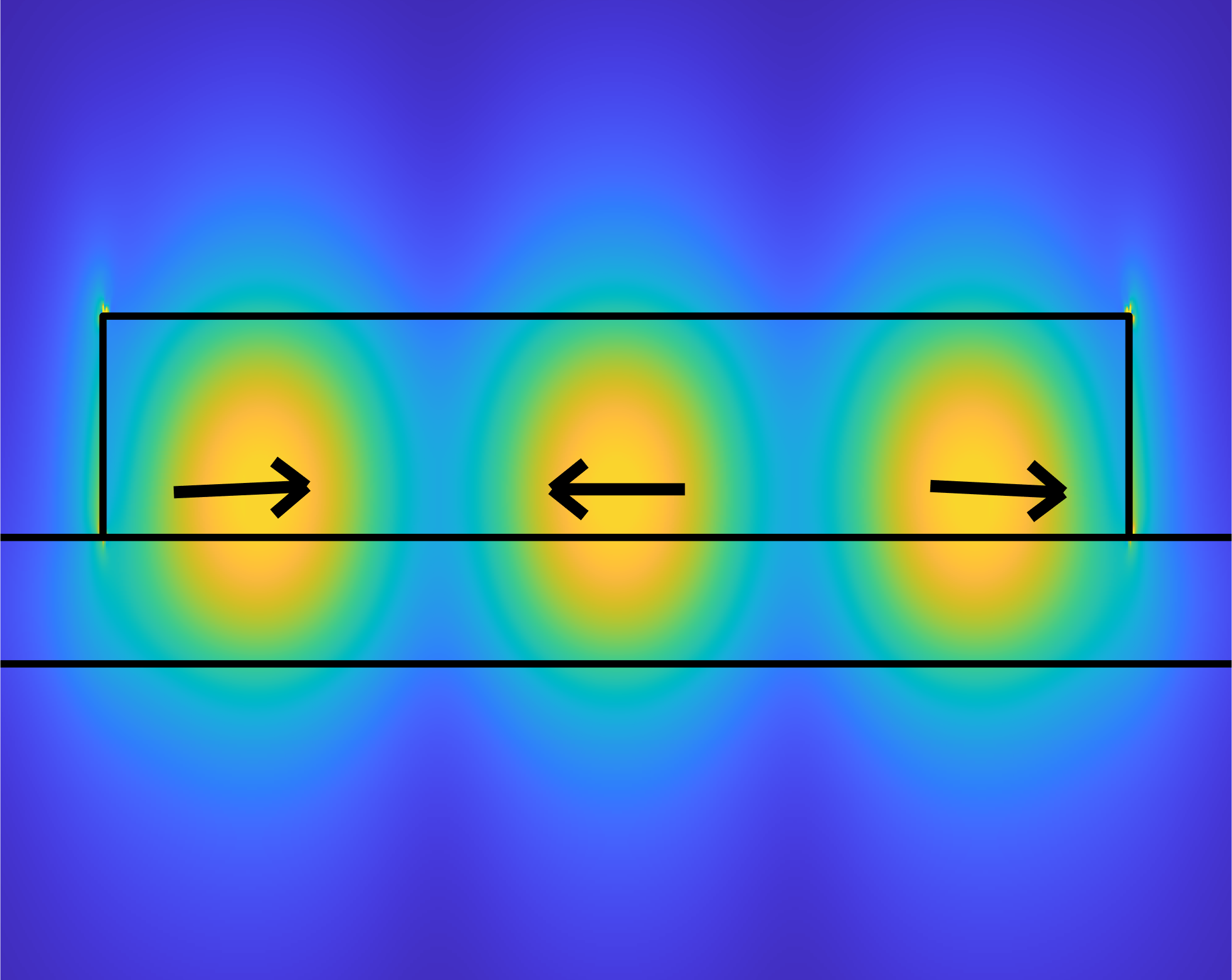};
\end{axis}
\begin{axis}[
    name=plot_2_4,
    at = (plot_2_3.east),
    anchor = west,
    xshift = 0.2cm,
    width=0.24\figW,
    height=0.24\figW*0.8,
    scale only axis,
    axis on top,
    ticks=none,
    enlargelimits=false,
]
\addplot graphics [xmin=-1168.422,xmax=1168.422,ymin=-200,ymax=420] {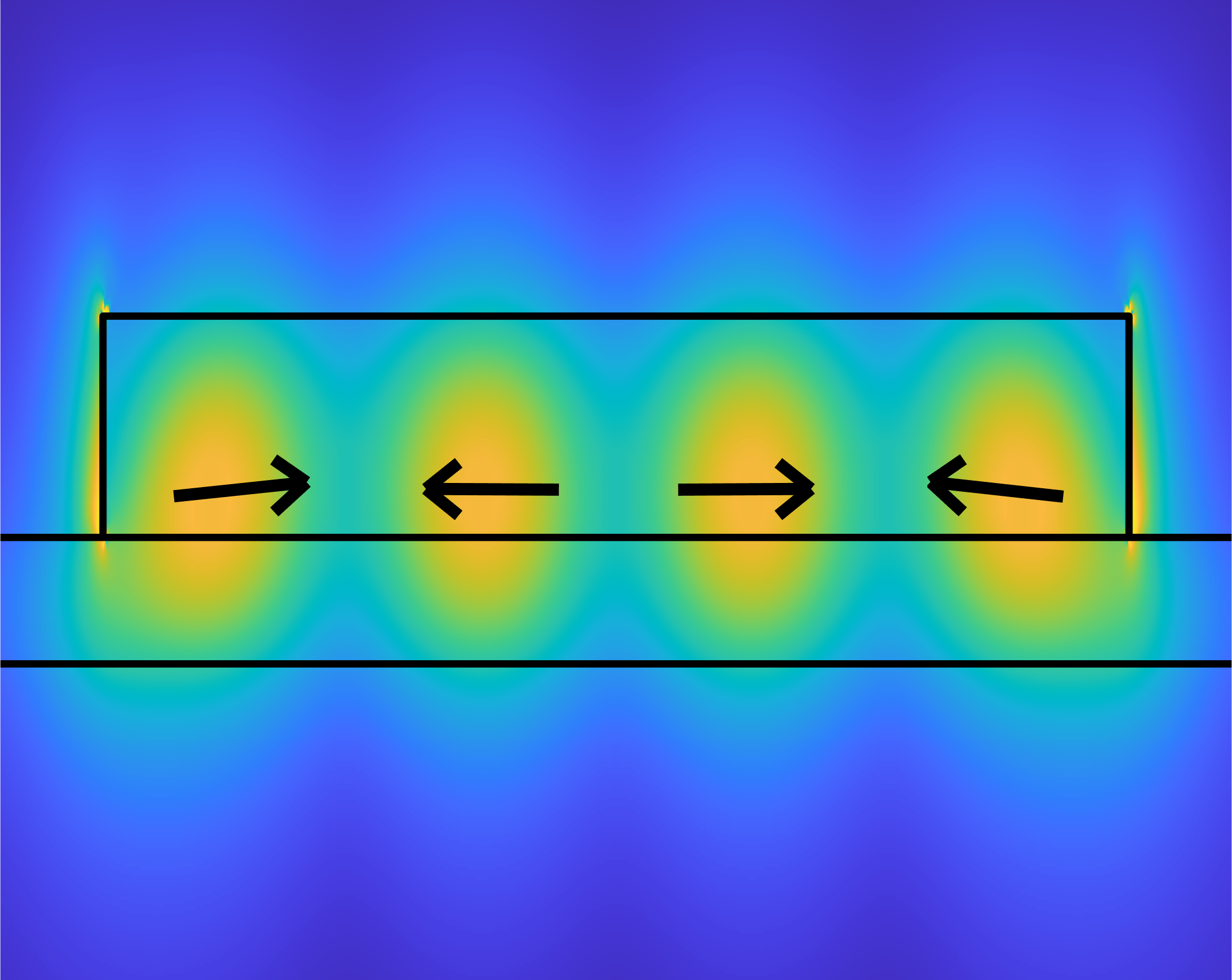};
\end{axis}
\end{tikzpicture}
    \caption{
        Transversal electrical field distributions (magnitudes and polarization directions) of some \cgls{fmf} and \cgls{nr} waveguide modes.\\
        \cgls{fmf} modes in the upper row: \HE{11e}, \TE{01}, \EH{11o} and \HE{12o}.\\
        \cgls{nr}  modes in the lower row: \TE{0}, \TE{1}, \TE{2} and \TE{3}.
    }
    \label{fig:mf_fmf_nr}
\end{figure}
\begin{figure}[tb]
    \centering
    \footnotesize
    \setlength\figW{0.7\linewidth} \input{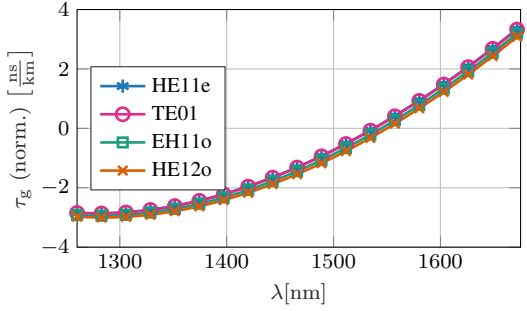}
    \caption{Relative normalized group delay of modes in a \cgls{fmf} with $r_\text{Core} = \qty{22}{\um}$, $w_\text{Trench} = \qty{0}{\um}$ and $r_\text{Clad} = \qty{32}{\um}$.}
    \label{fig:disp_fmf}
\end{figure}
\begin{figure}[tb]
    \centering
    \footnotesize
    \setlength\figW{0.7\linewidth} \input{figs/disp_nr}
    \caption{
        Relative normalized group delay of modes in a \cgls{nr} waveguide with
        $w_\text{Rib} = \qty{1947}{\nm}$ and $h_\text{Slab} = \qty{80}{\nm}$.
    }
    \label{fig:disp_nr}
\end{figure}
\begin{figure}[tb]
    \centering
    \footnotesize
    \setlength\figW{0.9\linewidth} \begin{tikzpicture}
\definecolor{colorSurf}{rgb}{1,0.8824,0.1098}%
\definecolor{colorSP1}{rgb}{0.90588,0.16078,0.54118}%
\definecolor{colorP2P1}{rgb}{0.10588,0.61961,0.46667}%
\definecolor{colorBestPM}{rgb}{0.85098,0.37255,0.00784}%
\begin{axis}[
    width=0.72\figW,
    height=0.4\figW,
    enlargelimits=false, 
    scale only axis,
    axis on top,
    xlabel={$f_{\mathrm{S}} \, \mathrm{\left[THz\right]}$},
    ylabel={$f_{\mathrm{P_2}} \, \mathrm{\left[THz\right]}$},
    xlabel style={font=\color{white!15!black}},
    ylabel style={font=\color{white!15!black}, yshift=-1mm},
    xtick pos=bottom, 
    colorbar,
    point meta min=0, point meta max=1,
    colorbar style={
        ylabel={FWM Efficiency},
        at={(1.03,0.5)},
        anchor=west,
        ytick style={color=black},
        tick pos=right,
        width=8,
    },
    legend style={at={(0.01,0.5)}, anchor=west, legend cell align=left, align=left, font=\tiny},
]
\addplot [color=colorSurf,line width=1.0pt] graphics [xmin=215,xmax=237.931,ymin=215,ymax=237.931] {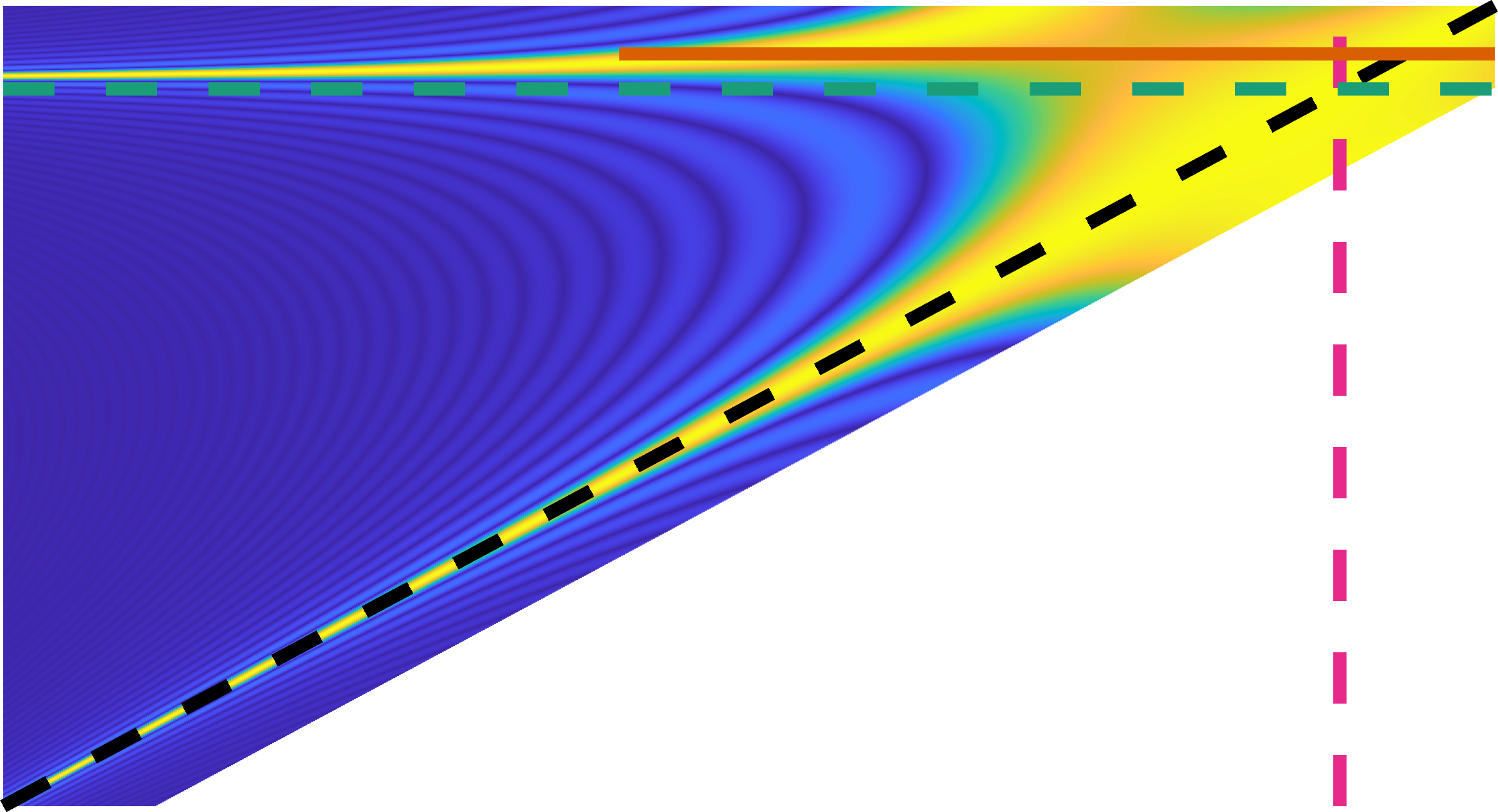};
\addlegendentry{FWM Efficiency}

\addplot [color=black,      line width=1.0pt, dashed] coordinates { (220,220) };  \addlegendentry{$f_{\mathrm{S}}   = f_{\mathrm{P_2}}$}
\addplot [color=colorSP1,   line width=1.0pt, dashed] coordinates { (220,220) };  \addlegendentry{$f_{\mathrm{S}}   = f_{\mathrm{P_1}}$}
\addplot [color=colorP2P1,  line width=1.0pt, dashed] coordinates { (220,220) };  \addlegendentry{$f_{\mathrm{P_2}} = f_{\mathrm{P_1}}$}
\addplot [color=colorBestPM,line width=1.0pt        ] coordinates { (220,220) };  \addlegendentry{\rev{Maximal $\qfwmBw$}}
\end{axis}
\begin{axis}[%
    width=0.72\figW,
    height=0.4\figW,
    scale only axis,
    axis on top,
    x dir=reverse,
    xmin=1260,
    xmax=1394.38,
    domain=1260:1394.38,
    xtick pos=top,
    xlabel={$\lambda_{\mathrm{S}}   \, \mathrm{\left[nm\right]}$},
    xlabel style={font=\color{white!15!black}},
    axis x line*=top,
    axis y line=none,
]
\addplot[draw=none,fill=none] {x}; 
\end{axis}
\end{tikzpicture}
    \caption{
        \rev{Normalized \cgls{fwm} efficiency for a fiber with $r_\text{Core} = \qty{22}{\um}$, $w_\text{Trench} = \qty{0}{\um}$ and $r_\text{Clad} = \qty{32}{\um}$.
        Here, $\mathrm{S}$ and $\mathrm{P}_2$  propagate in \HE{11e} and $\mathrm{P}_1$ and $\mathrm{I}$ in \HE{21e}, and pump 1 is fixed at $f_{\mathrm{P}_1} = \qty{235.55}{\THz}$ ($\qty{1272.7}{\nm}$).}
        For combinations of $f_\mathrm{S}$ and $f_{\mathrm{P}_2}$ in white areas, the idler frequency $f_\mathrm{I}$ lies outside of the simulated frequency range (O- to U-band).
    }
    \label{fig:pm_fmf}
\end{figure}
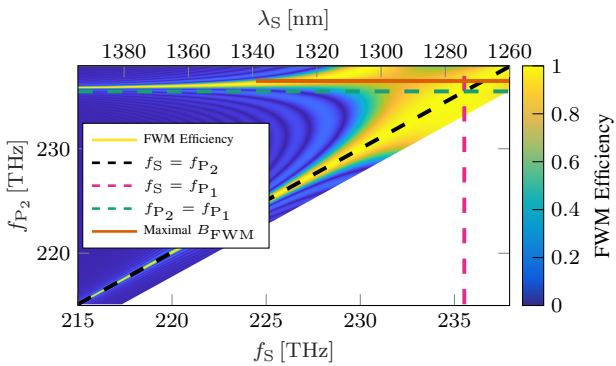
In this section, we present refractive index profiles, mode fields, and \rev{group delay} curves of both, optical fibers and \cgls{nr} silicon waveguides.
\CrefSubfig{fig:n_fmf_nr}{a} shows the \cgls{fmf} geometry with core radius $r_\text{Core}$, cladding radius $r_\text{Clad}$ and trench width $w_\text{Trench}$.
We use graded index fibers with a depressed cladding, c.f. \cite{maruyamaTwoModeOptical2014}.
Note that the refractive index contrast is very low, which is typical for optical fibers.
The index profile has a grading exponent of $2.0$ and an ellipticity of $1.0$ (i.e., perfectly round).
Similarly, \crefSubfig{fig:n_fmf_nr}{b} shows the \cgls{nr} waveguide geometry with rib width $w_\text{Rib}$, slab height $h_\text{Slab}$, and \cgls{soi} height $h_\text{\acrshort{soi}}$. The waveguide consists of a crystalline silicon core surrounded by silica, which results in a very high refractive index contrast.
\Cref{tab:simParams} lists the geometry values we used for optimizations.

We use a full vectorial \cgls{fdm} mode solver based on \cite{fallahkhairVectorFiniteDifference2008} to compute waveguide modes and propagation constants, from which we can derive \rev{normalized} group delay and chromatic dispersion curves

\begin{align}
    \tau_g^\mA(\lambda) = \dd{\beta^\mA}{\omega},\qquad
    D^\mA(\lambda) = \dd{\tau_g^\mA}{\lambda}.\label{eq:groupDelayAndDispersion}
\end{align}
\rev{The normalized group delay $\tau_g$ (also called inverse group velocity) is the time per distance a signal accumulates while propagating in a waveguide.
It is common to present relative values, which simply means that all curves are shifted by a constant offset -- such that the slowest mode has value zero at some arbitrary wavelength (typically \qty{1550}{\nm}).}
\rev{To improve simulation accuracy, we don't fix the cladding radius to one value, but let it vary with the core radius.
This way, more points of the mode solver's discretization grid are available for computing the fields in the core.
We found that \qty{10}{\um} outside of the core, the fields have decayed enough to be negligible.}
\Cref{fig:mf_fmf_nr} shows the four lowest-order (largest $\beta$) computed vectorial mode fields for both types of waveguides.
Since fiber modes appear in groups with (almost) identical propagation constants (e.g., \{\HE{11e}, \HE{11o}\} or \{\TE{01}, \TM{01}, \HE{21e}, \HE{21o}\}), we show one mode of the four lowest order \textit{mode groups} instead.
While the fiber modes have both transversal components, the \acrlong{nr} modes are approximately linearly polarized.
Note that the axes have different scalings.

The group delays for one exemplary \cgls{fmf} and \cgls{nr} waveguide are shown in \cref{fig:disp_fmf,fig:disp_nr}, respectively.
Note that the group delays of the two waveguide types differ by orders of magnitude and also in shape (especially the weakly guided \TE{3} mode).

Our waveguide optimization relies on the \cgls{fwm} efficiency
\begin{align}
    \qfwmEff(\Delta\beta) = \frac{ 1 - \exps{-(\alpha + j \Delta\beta) L} }{(\alpha + j \Delta\beta) L}\label{eq:eta}
\end{align}
with waveguide length $L$ and attenuation $\alpha$.
\rev{It was first derived in \cite{hillCWThreewaveMixing1978} and extended to \cglspl{fmf} in \cite{rademacherInvestigationIntermodalFourwave2018}, by using \namecref{eq:eta}~(17) from \cite{xiaoTheoryIntermodalFourwave2014}.
A very similar idea is used, e.g., in \cite{borghiNonlinearSiliconPhotonics2017} to assess \cgls{fwm} efficiency in silicon rib waveguides.
Note that this is an approximative equation for a best-case analysis and we highlight some drawbacks in \cref{sec:propa}.}

\rev{We always used $L = \qty{2}{\cm}$ and $\alpha = \qty{1}{\dB\per\cm}$ for \cgls{nr} waveguides and $L = \qty{10}{\m}$ (see \cref{sec:34PM} why) and $\alpha = \qty{0.226}{\dB\per\km}$ for fibers.
We always jointly optimize laser wavelengths (signal, pumps and idler) and the choice of modes, with the goal of maximal $\qfwmBw$.}

\rev{\Cref{fig:pm_fmf} shows an example of the normalized \cgls{fwm} efficiency $\abs{\qfwmEff} / \max(\abs{\qfwmEff})$ for a \cgls{fmf} with $r_\text{Core} = \qty{22}{\um}$, $w_\text{Trench} = \qty{0}{\um}$ and $r_\text{Clad} = \qty{32}{\um}$.
The \cgls{fwm} efficiency is shown as a function of pump 2 and signal frequency and all other parameters were optimized for maximal \cgls{fwm} bandwidth $\qfwmBw$ in a \cgls{2fwm} operation mode (it was enforced that two modes are used with two lasers in each).
The optimal values found by the optimization are $\mathrm{S}$ and $\mathrm{P}_2$ in \HE{11e} and $\mathrm{P}_1$ and $\mathrm{I}$ in \HE{21e}, and pump 1 fixed at $f_{\mathrm{P}_1} = \qty{235.55}{\THz}$ (or $\lambda_{\mathrm{P}_1} = \qty{1272.7}{\nm}$), leading to the presented plot and the marked $\qfwmBw$.
It is achieved for $f_{\mathrm{P}_2} = \qty{236.6}{\THz}$ (or $\lambda_{\mathrm{P}_2} = \qty{1267.1}{\nm}$).
For this choice, \cgls{pm} is retained for signal frequencies between \qtyrange{224.93}{237.93}{\THz} (\qtyrange{1260.0}{1332.8}{\nm}). }
The dashed lines mark where two lasers have the same frequency and hence \gls{fwm} is degenerate.
\rev{To avoid degenerate \cgls{fwm}, we remove configurations from the search space where signal or idler are closer to a pump than \qty{250}{\GHz} (roughly \qty{2}{\nm}).
Separating the idler from pumps is strictly necessary, since waveguide imperfections always couple the strong pumps to the idler's mode as well.
Thus, it is impossible to separate the idler at the end of the waveguide.
Enforcing a separation between the two pumps is not strictly necessary, but in cases where they are not, the \cgls{bs} idler has the same frequency as the input signal (see \cref{eq:fwmEnergyConservationBs}), which, of course, disables wavelength conversion.
Allowing the signal to have one of the pump's frequency has a similar effect: the \cgls{opc} idler is generated at the other pump's frequency (see \cref{eq:fwmEnergyConservationPc}) and can't be separated any more.
For further optimization results for \cgls{nr} waveguides, we refer to \cite{kernetzkyMultiDimensionalOptimization2020}.}

For the results in this \rev{feasibility} study, we allow all lasers to be in the O-, E-, S-, C-, L or U-bands (\qtyrange{1260}{1675}{\nm} or \qtyrange{179}{238}{\tera\Hz}) without any limitation and don't restrict \cgls{fwm} to \textit{useful} configurations (e.g., wavelength conversion from C- to O-band, etc.) -- in contrast to \cite{kernetzkyMultiDimensionalOptimization2020}.

\section{\rev{FWM Bandwidth Optimization Results}}\label{sec:34PM}
\begin{figure*}[tb]
    \centering
    \footnotesize
    \setlength\figW{0.8\linewidth} \begin{tikzpicture}
\begin{axis}[
    name = resMuFmfOnePic,
    enlargelimits=false,
    axis on top,
    width=0.72\figW/2,
    height=0.35\figW/2,
    scale only axis,
    xticklabels={},
    ylabel={$w_\text{Trench} \, \mathrm{\left[\mu{}m\right]}$},
    ylabel style={font=\color{white!15!black}, yshift=-1mm},
    ytick={0, 1, 2, 4, 8},
    yticklabel style={text width=2em, align=right}, 
    point meta min=1.34861, point meta max=3.13553,
    colorbar,
    colorbar style={
        ylabel={FWM BW $\qfwmBw$},
        at={(1.03,0.5)},
        anchor=west,
        ytick style={color=black},
        tick pos=right,
        width=8,
    },
]
\addplot graphics [xmin=4.9928,xmax=41.0072,ymin=-0.12665,ymax=8.12665] {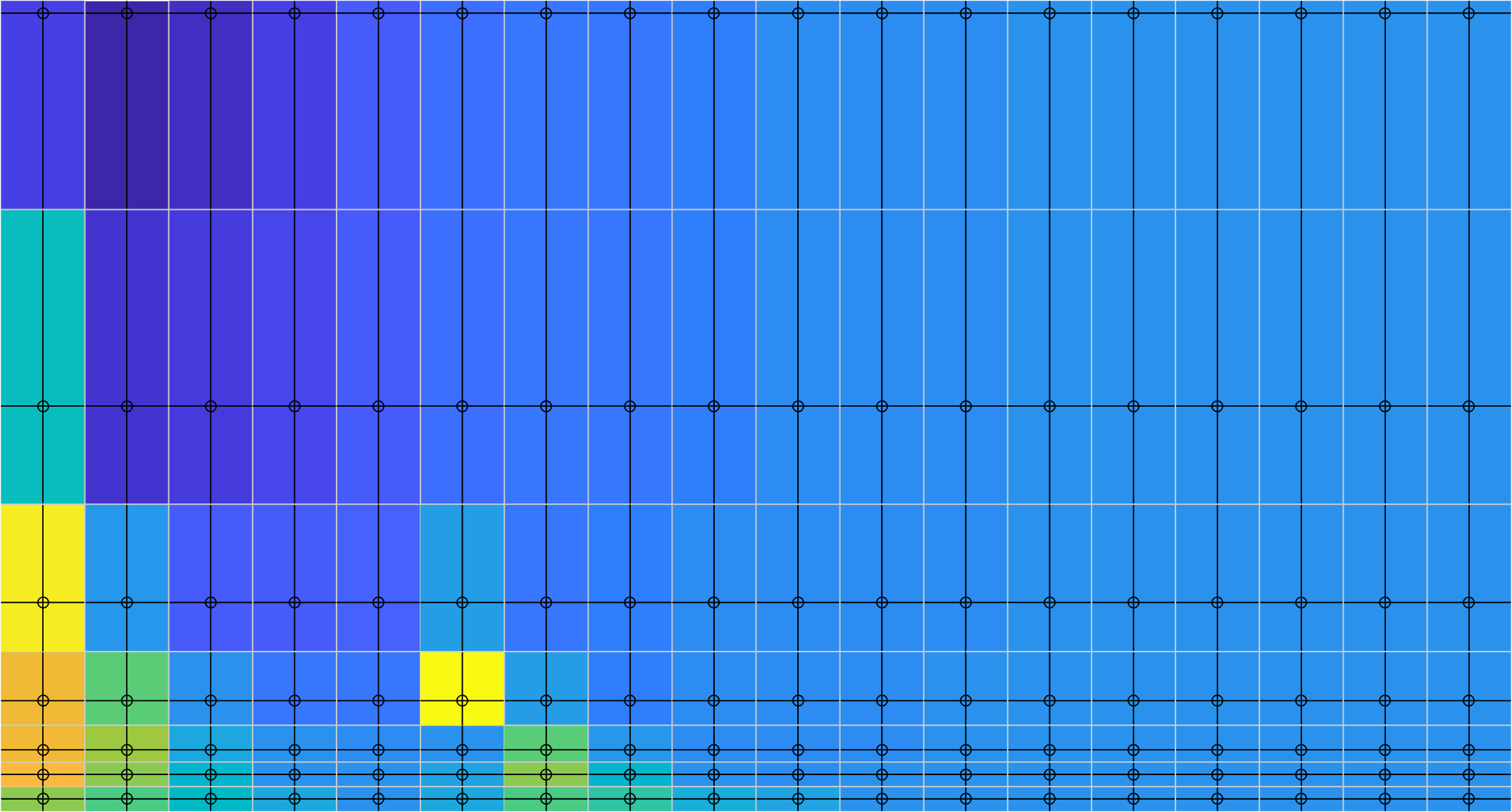};
\end{axis}
\begin{axis}[
    name = resMuFmfTwoPic,
    at = (resMuFmfOnePic.south),
    anchor = north,
    yshift = -0.3cm,
    enlargelimits=false,
    axis on top,
    width=0.72\figW/2,
    height=0.35\figW/2,
    scale only axis,
    xticklabels={},
    ylabel={$w_\text{Trench} \, \mathrm{\left[\mu{}m\right]}$},
    ylabel style={font=\color{white!15!black}, yshift=-1mm},
    ytick={0, 1, 2, 4, 8},
    yticklabel style={text width=2em, align=right}, 
    point meta min=1.51719, point meta max=3.10181,
    colorbar,
    colorbar style={
        ylabel={FWM BW $\qfwmBw$},
        at={(1.03,0.5)},
        anchor=west,
        ytick style={color=black},
        tick pos=right,
        width=8,
    },
]
\addplot graphics [xmin=4.9928,xmax=41.0072,ymin=-0.12665,ymax=8.12665] {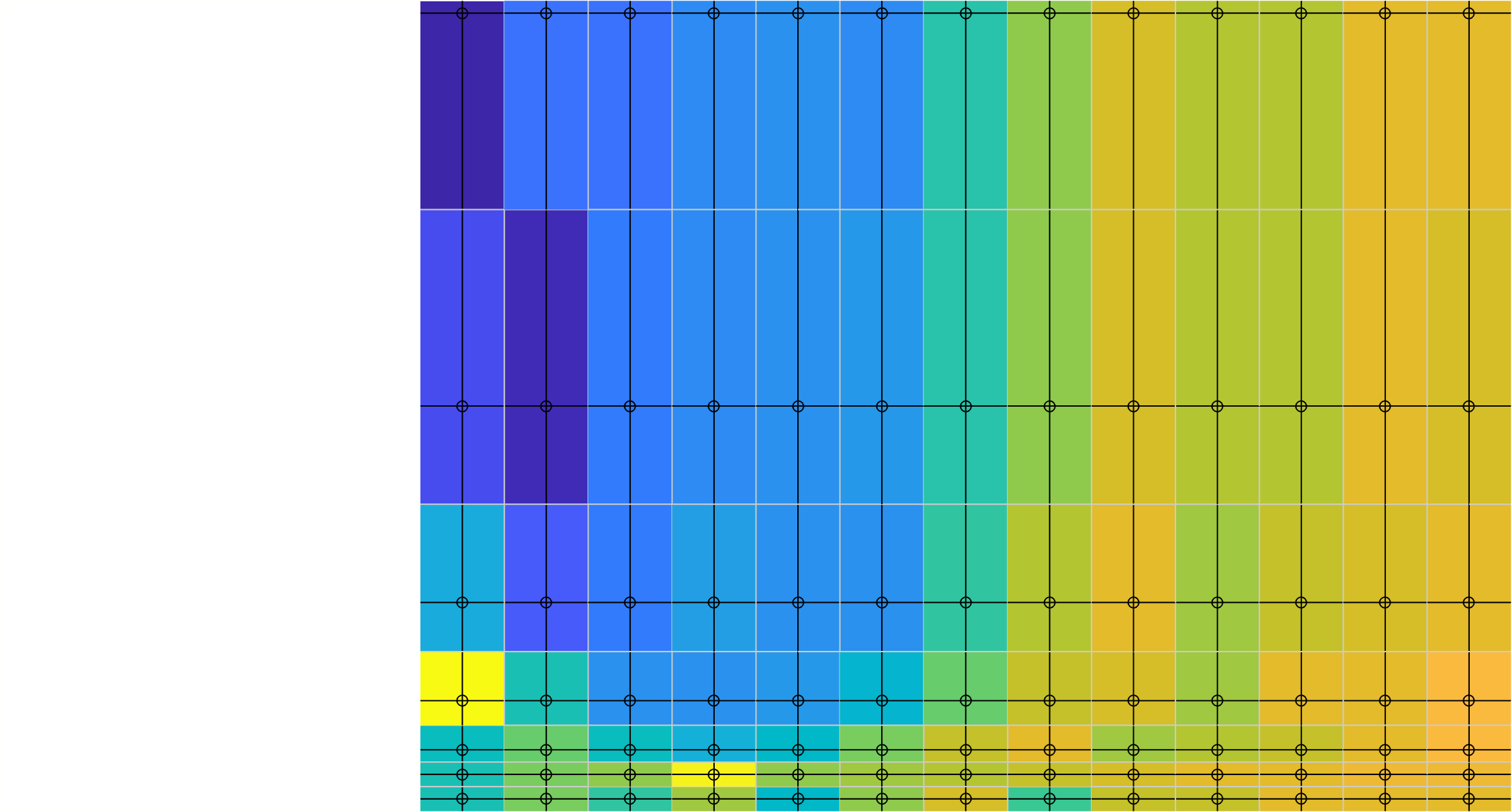};
\end{axis}
\begin{axis}[
    name = resMuFmfThreePic,
    at = (resMuFmfTwoPic.south),
    anchor = north,
    yshift = -0.3cm,
    enlargelimits=false,
    axis on top,
    width=0.72\figW/2,
    height=0.35\figW/2,
    scale only axis,
    xlabel style={font=\color{white!15!black}},
    xlabel={$r_\text{Core} \, \mathrm{\left[\mu{}m\right]}$},
    ylabel={$w_\text{Trench} \, \mathrm{\left[\mu{}m\right]}$},
    ylabel style={font=\color{white!15!black}, yshift=-1mm},
    ytick={0, 1, 2, 4, 8},
    yticklabel style={text width=2em, align=right}, 
    point meta min=0.0337153, point meta max=3.10181,
    colorbar,
    colorbar style={
        ylabel={FWM BW $\qfwmBw$},
        at={(1.03,0.5)},
        anchor=west,
        ytick style={color=black},
        tick pos=right,
        width=8,
    },
]
\addplot graphics [xmin=4.9928,xmax=41.0072,ymin=-0.12665,ymax=8.12665] {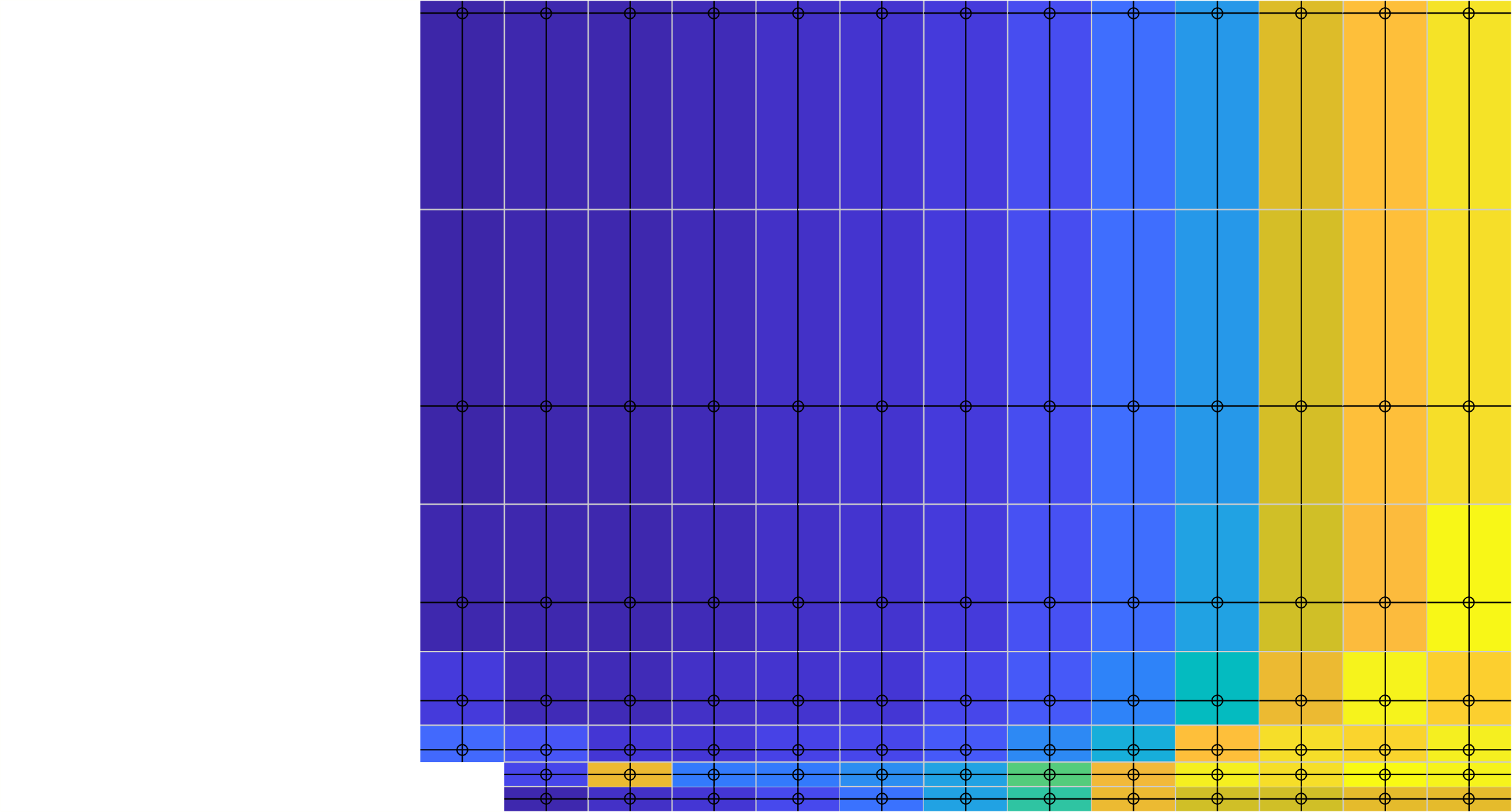};
\end{axis}
\begin{axis}[
    name = resMuNrOnePic,
    at = (resMuFmfOnePic.east),
    anchor = west,
    xshift = 3cm,
    enlargelimits=false,
    axis on top,
    width=0.72\figW/2,
    height=0.35\figW/2,
    scale only axis,
    xticklabels={},
    ylabel={$h_\text{Slab} \, \mathrm{\left[nm\right]}$},
    ylabel style={font=\color{white!15!black}, yshift=-1mm},
    ytick={70, 90, 110, 130, 150, 170},
    yticklabel style={text width=2em, align=right}, 
    colorbar,
    point meta min=0.370869, point meta max=12.1038,
    colorbar style={
        ylabel={FWM BW $\qfwmBw$},
        at={(1.03,0.5)},
        anchor=west,
        ytick style={color=black},
        tick pos=right,
        width=8,
    },
]
\addplot graphics [xmin=947.368,xmax=3052.63,ymin=64.976,ymax=185.024] {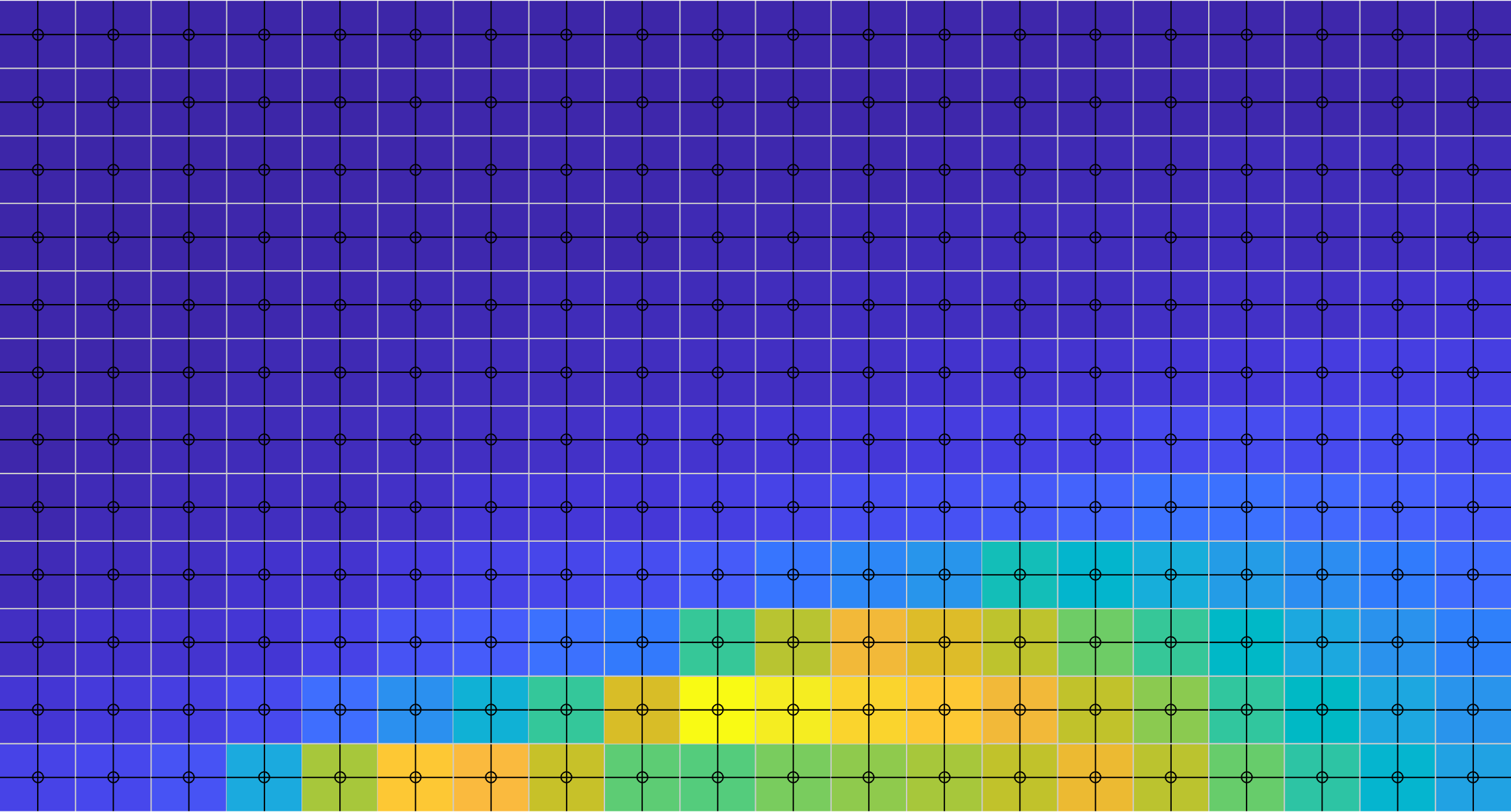};
\end{axis}
\begin{axis}[
    name = resMuNrTwoPic,
    at = (resMuNrOnePic.south),
    anchor = north,
    yshift = -0.3cm,
    enlargelimits=false,
    axis on top,
    width=0.72\figW/2,
    height=0.35\figW/2,
    scale only axis,
    xticklabels={},
    ylabel={$h_\text{Slab} \, \mathrm{\left[nm\right]}$},
    ylabel style={font=\color{white!15!black}, yshift=-1mm},
    ytick={70, 90, 110, 130, 150, 170},
    yticklabel style={text width=2em, align=right}, 
    colorbar,
    point meta min=0, point meta max=9.4403,
    colorbar style={
        ylabel={FWM BW $\qfwmBw$},
        at={(1.03,0.5)},
        anchor=west,
        ytick style={color=black},
        tick pos=right,
        width=8,
    },
]
\addplot graphics [xmin=947.368,xmax=3052.63,ymin=64.976,ymax=185.024] {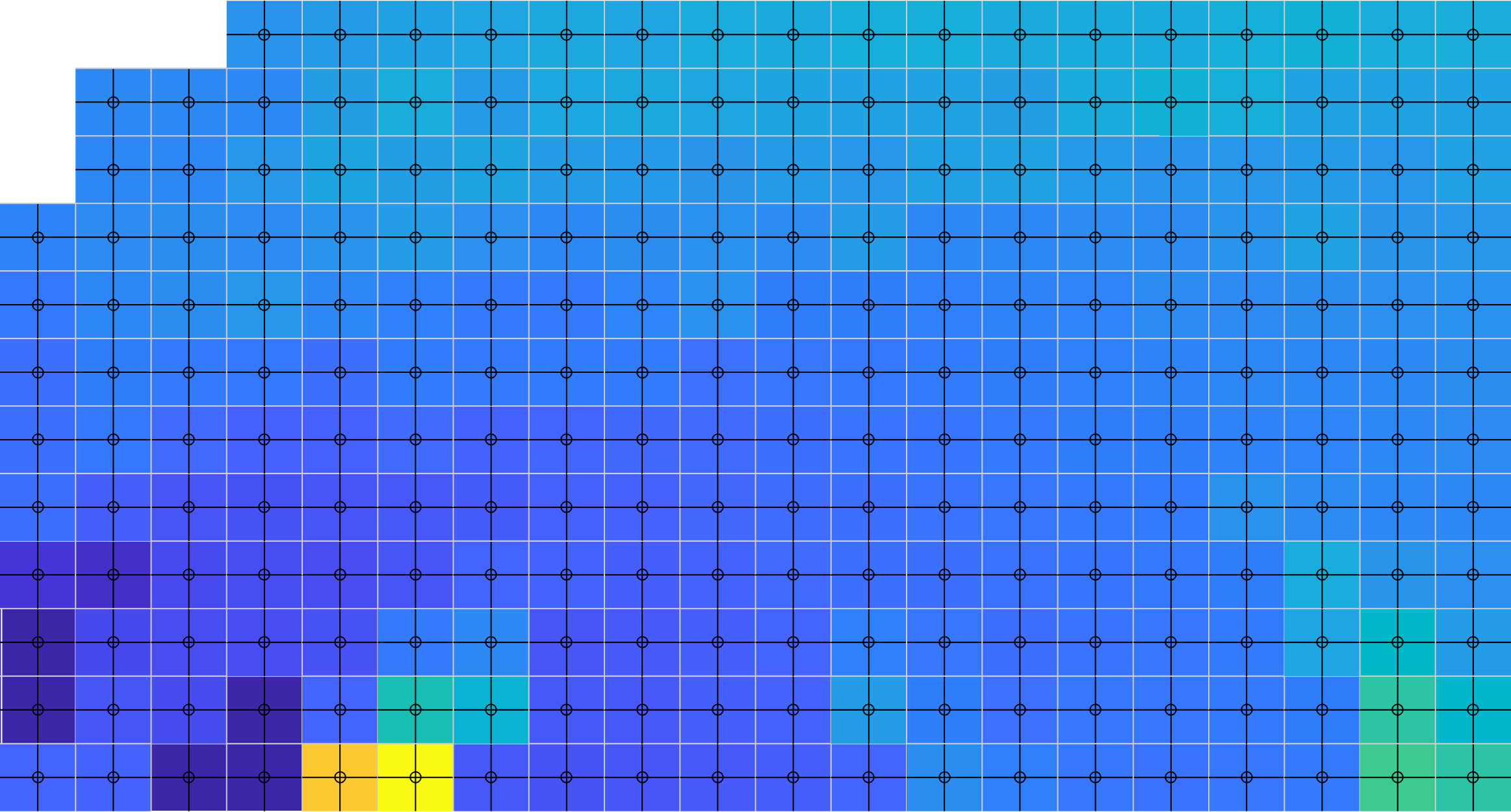};
\end{axis}
\begin{axis}[
    name = resMuNrThreePic,
    at = (resMuNrTwoPic.south),
    anchor = north,
    yshift = -0.3cm,
    enlargelimits=false,
    axis on top,
    width=0.72\figW/2,
    height=0.35\figW/2,
    scale only axis,
    xlabel style={font=\color{white!15!black}},
    xlabel={$w_\text{Rib} \, \mathrm{\left[nm\right]}$},
    xtick={1000, 1421, 1842, 2157, 2578, 3000},
    ylabel={$h_\text{Slab} \, \mathrm{\left[nm\right]}$},
    ylabel style={font=\color{white!15!black}, yshift=-1mm},
    ytick={70, 90, 110, 130, 150, 170},
    yticklabel style={text width=2em, align=right}, 
    colorbar,
    point meta min=0, point meta max=1.55091,
    colorbar style={
        ylabel={FWM BW $\qfwmBw$},
        at={(1.03,0.5)},
        anchor=west,
        ytick style={color=black},
        tick pos=right,
        width=8,
    },
]
\addplot graphics [xmin=947.368,xmax=3052.63,ymin=64.976,ymax=185.024] {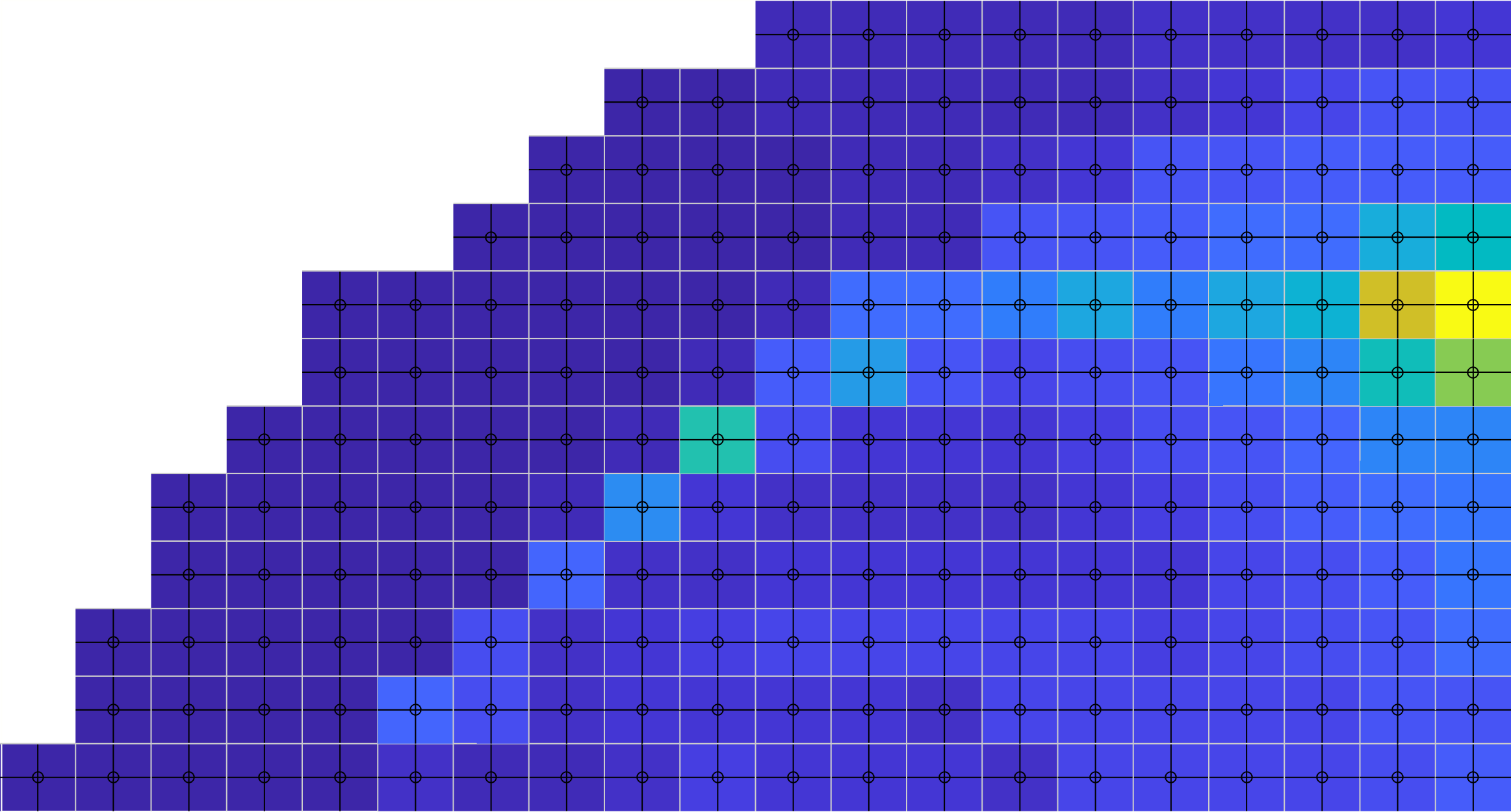};
\end{axis}
\node [left=of resMuFmfOnePic.south west,   anchor=south east, shift=({0.2, -0.4})] {(a)};
\node [left=of resMuFmfTwoPic.south west,   anchor=south east, shift=({0.2, -0.4})] {(b)};
\node [left=of resMuFmfThreePic.south west, anchor=south east, shift=({0.2, -0.4})] {(c)};
\node [left=of resMuNrOnePic.south west,    anchor=south east, shift=({0.2, -0.4})] {(d)};
\node [left=of resMuNrTwoPic.south west,    anchor=south east, shift=({0.2, -0.4})] {(e)};
\node [left=of resMuNrThreePic.south west,  anchor=south east, shift=({0.2, -0.4})] {(f)};
\end{tikzpicture}
    \caption{
        Highest achievable \cgls{fwm} bandwidths for \cglspl{fmf} and \cgls{nr} waveguides with different geometries.
        All normalized to multiples of the C-band's frequency range.
        Waveguides in white areas don't support enough guided modes.
        (a) \cGls{1fwm} in \cglspl{fmf}.
        (b) \cGls{2fwm} in \cglspl{fmf}.
        (c) \cGls{3fwm} in \cglspl{fmf}.
        (d) \cGls{1fwm} in \cgls{nr} waveguides.
        (e) \cGls{2fwm} in \cgls{nr} waveguides.
        (f) \cGls{3fwm} in \cgls{nr} waveguides.
    }
    \label{fig:res_regular_mu}
\end{figure*}
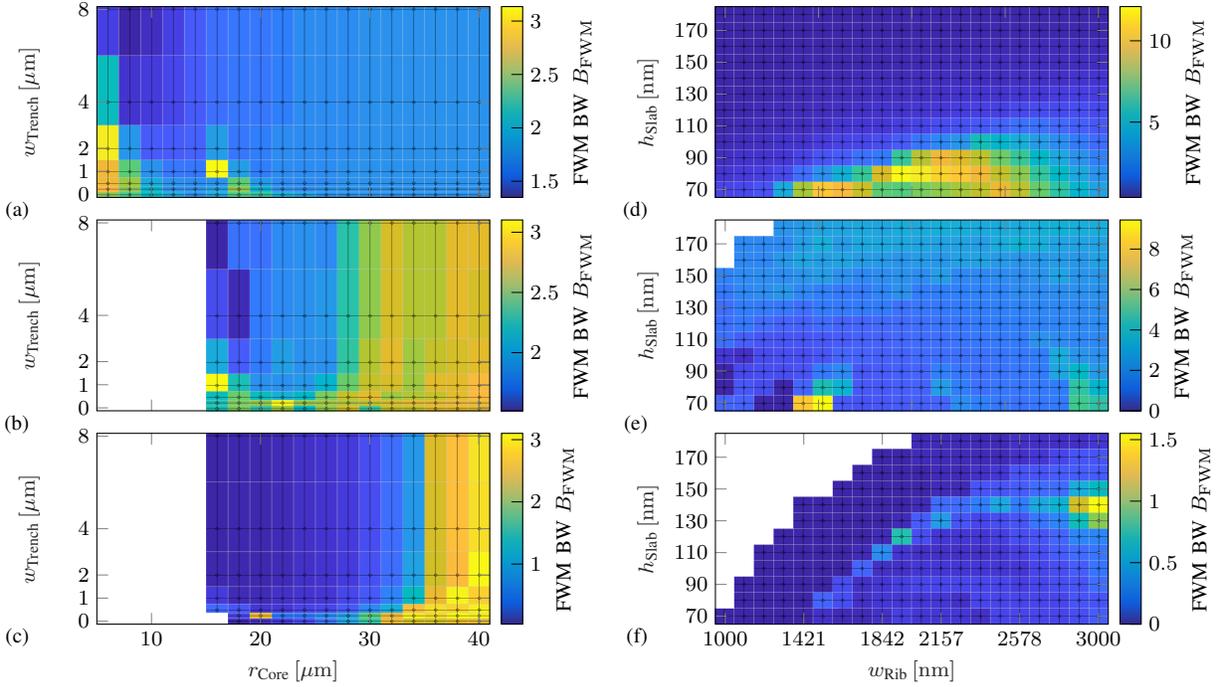
\begin{figure}[tb]
    \centering
    \footnotesize
    \setlength\figW{0.7\linewidth} \input{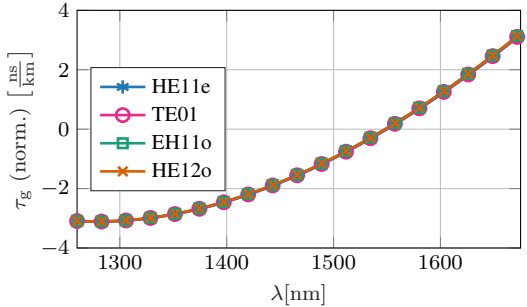}
    \caption{Relative normalized group delay of modes in a \cgls{fmf} with $r_\text{Core} = \qty{40}{\um}$, $w_\text{Trench} = \qty{0}{\um}$ and $r_\text{Clad} = \qty{50}{\um}$.}
    \label{fig:disp_fmf_large_core}
\end{figure}
Repeating the search for highest \cgls{fwm} bandwidth like in \cref{fig:pm_fmf} for different waveguides (we used the parameters listed in \cref{tab:simParams}), allows to compare waveguide geometries.
We performed \rev{separate} waveguide parameter optimizations for 1-, 2-, 3- and \cgls{4fwm}.
\Cref{fig:res_regular_mu} shows the best achievable \cgls{fwm} bandwidth $\qfwmBw$ for \cgls{fmf} and \cgls{nr} waveguides.
The left subfigures (a) to (c) show results for \cglspl{fmf} and subfigures (d) to (f) for \cgls{nr} waveguides.
The first row shows \cgls{1fwm}, the second row \cgls{2fwm} and the third row \cgls{3fwm}.

Both waveguides have negligible \cgls{fwm} bandwidths for \cgls{4fwm}.
Therefore, we ignore \cgls{4fwm} in the rest of this paper with the conclusion that it is not feasible.

\rev{The optimizations for \cglspl{fmf} were performed with only \qty{10}{\m} fiber length to be able to cope with the large searched bandwidth (from O- to U-band).
With the reduced length, the \cgls{pm} bandwidth becomes larger and is better suited for numerical optimizations.}
In reality, the length needs to be in the range of several hundred meters up to some kilometers to build up sufficient idler powers.
This means that \cgls{fwm} bandwidths of \cglspl{fmf} are much lower in reality (see, e.g., \cite{rademacherInvestigationIntermodalFourwave2018} and note that length reduces bandwidth in \cref{eq:eta}).
Our results for \cglspl{fmf} are therefore not comparable to our results for \cgls{nr} waveguides.
This is not a problem, however, since our goal is to compare \cgls{fwm} with different numbers of modes, among the \textit{same type} of waveguides.

From \cref{fig:res_regular_mu} (a) to (c), one can conclude that \cgls{fwm} with one, two and three modes gives roughly the same maximal \cgls{fwm} bandwidth in \cglspl{fmf}.
However, the higher the number of modes, the more the optima are concentrated towards larger core radii.
To understand the reason, compare the fiber with $r_\text{Core} = \qty{22}{\um}$ in \cref{fig:disp_fmf} with a second fiber with $r_\text{Core} = \qty{40}{\um}$ in \cref{fig:disp_fmf_large_core}.
Due to the large core, all the considered mode groups have very similar group delays and the process effectively becomes \cgls{1fwm} for any number of used modes.

The two best waveguides in \cref{fig:res_regular_mu} (a) have broad zero dispersion regions of one mode in the considered frequency range, which leads to high \cgls{fwm} bandwidths (see \cref{eq:pmApproxOneMode}).

For \cgls{nr} waveguides, comparing \cref{fig:res_regular_mu} (d) to (f) reveals that increasing the number of modes decreases \cgls{fwm} bandwidths.
Here, larger cores also lead to approaching group delay curves (similar to \cref{fig:disp_fmf} vs. \cref{fig:disp_fmf_large_core}), but the difference stays much larger and the process does not effectively become \cgls{1fwm} for the geometries simulated here.
The difference in group delay curve shapes and magnitudes in \cref{fig:disp_fmf,fig:disp_nr} also indicates that \cgls{nr} waveguides behave differently.
By comparing the plots in \cref{fig:res_regular_mu}, it is clear that \cgls{fwm} in \cgls{nr} waveguides is more sensitive to geometry variations than in \cglspl{fmf}.
The ultra high bandwidths (more than 10 times the C-band) in \crefSubfig{fig:res_regular_mu}{d} are also based on low dispersion regions, see the \TE{3} mode in \cref{fig:disp_nr}.

\rev{The bandwidths for \cgls{opc} are very similar to \cref{fig:res_regular_mu}, we list the maxima:
(a): \num{5}$\mathrm{B}_\mathrm{C}$, (b): \num{1.8}$\mathrm{B}_\mathrm{C}$, (c): \num{1.8}$\mathrm{B}_\mathrm{C}$,
(d): \num{11}$\mathrm{B}_\mathrm{C}$, (e): \num{1}$\mathrm{B}_\mathrm{C}$, (f): \num{1}$\mathrm{B}_\mathrm{C}$, where $\mathrm{B}_\mathrm{C}$ is the C-band's width.}

\section{Considering Modal Overlap}\label{sec:overlap}
\begin{figure}[tb]
    \centering
    \footnotesize
    \setlength\figW{0.8\linewidth} \input{figs/ol}
    \caption{
        Examples of overlapped \cgls{nr} waveguide \rev{($w_\text{Rib} = \qty{1947}{\nm}$ and $h_\text{slab} = \qty{80}{\nm}$)} mode fields \mA{}, \mB{}, \mC{}, and \mD{} for two configurations.
        The fields are scaled for the plots and magnitude units are irrelevant here.\\
        (a): Acceptable nonlinear interaction with $\TE{3}, \TE{1}, \TE{2}, \TE{2}$.\\
        (b): Poor       nonlinear interaction with $\TE{3}, \TE{0}, \TE{2}, \TE{2}$.
    }
    \label{fig:ol}
\end{figure}
\begin{figure}[tb]
    \centering
    \footnotesize
    \setlength\figW{0.8\linewidth} \begin{tikzpicture}
\begin{axis}[
    name = resFmfTwoAbsGammaMuPic,
    enlargelimits=false,
    axis on top,
    width=0.72\figW,
    height=0.35\figW,
    scale only axis,
    xlabel style={font=\color{white!15!black}},
    xlabel={$r_\text{Core} \, \mathrm{\left[\mu{}m\right]}$},
    ylabel={$w_\text{Trench} \, \mathrm{\left[\mu{}m\right]}$},
    ylabel style={font=\color{white!15!black}, yshift=-1mm},
    ytick={0, 1, 2, 4, 8},
    yticklabel style={text width=2em, align=right}, 
    point meta min=0, point meta max=3.10181,
    colorbar,
    colorbar style={
        ylabel={FWM BW $\qfwmBw$},
        at={(1.03,0.5)},
        anchor=west,
        yticklabel style={/pgf/number format/fixed,},
        scaled y ticks=false,
        ytick style={color=black},
        tick pos=right,
        width=8,
    },
]
\addplot graphics [xmin=4.9928,xmax=41.0072,ymin=-0.12665,ymax=8.12665] {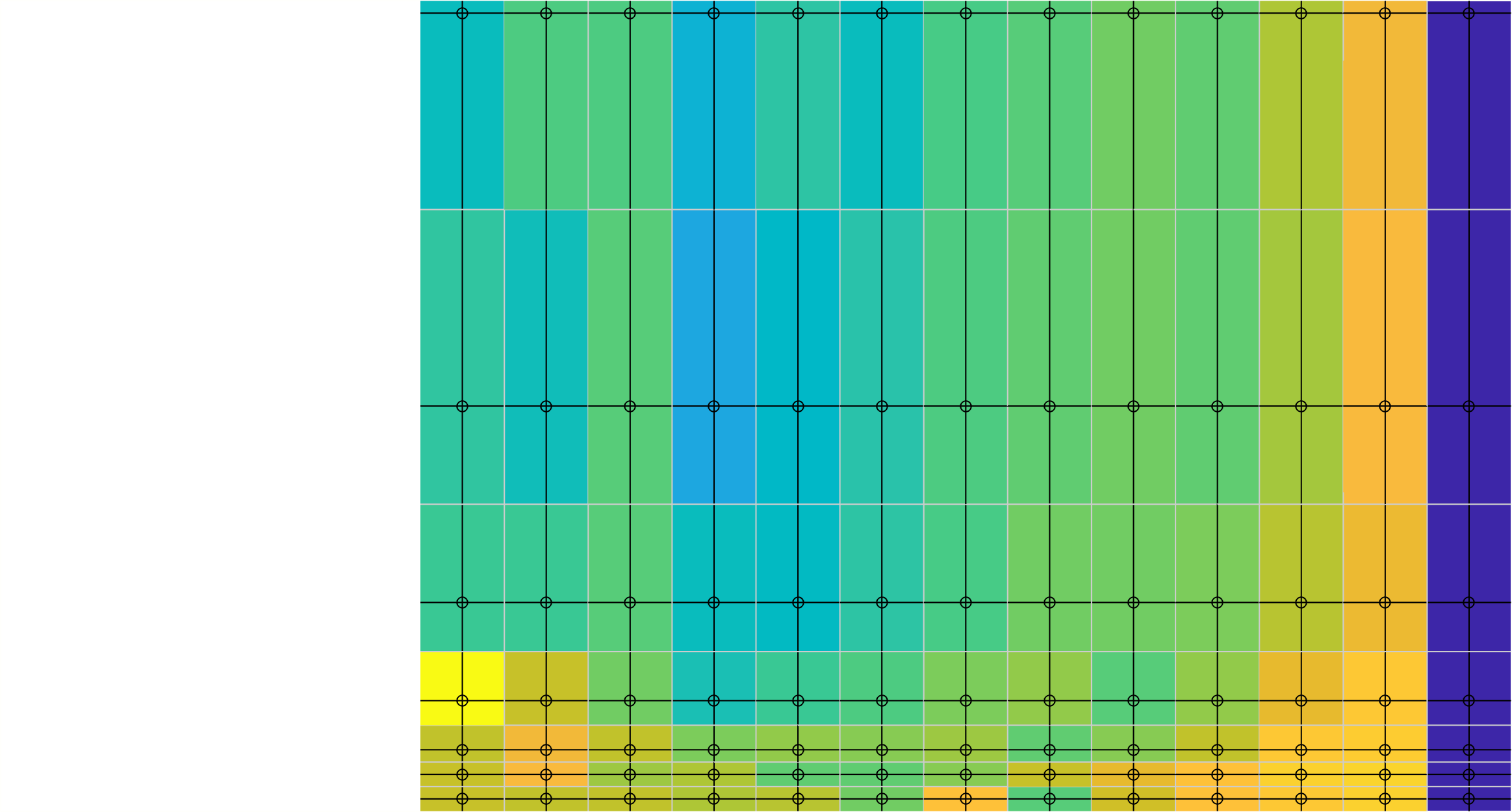};
\end{axis}
\begin{axis}[
    name = resNrThreeAbsGammaMuPic,
    at = (resFmfTwoAbsGammaMuPic.south),
    anchor = north,
    yshift = -1cm,
    enlargelimits=false,
    axis on top,
    width=0.72\figW,
    height=0.35\figW,
    scale only axis,
    xlabel style={font=\color{white!15!black}},
    xlabel={$w_\text{Rib} \, \mathrm{\left[nm\right]}$},
    xtick={1000, 1421, 1842, 2157, 2578, 3000},
    ylabel={$h_\text{Slab} \, \mathrm{\left[nm\right]}$},
    ylabel style={font=\color{white!15!black}, yshift=-1mm},
    ytick={70, 90, 110, 130, 150, 170},
    yticklabel style={text width=2em, align=right}, 
    point meta min=0, point meta max=0.0674307,
    colorbar,
    colorbar style={
        ylabel={FWM BW $\qfwmBw$},
        at={(1.03,0.5)},
        anchor=west,
        yticklabel style={/pgf/number format/fixed,},
        scaled y ticks=false,
        ytick style={color=black},
        tick pos=right,
        width=8,
    },
]
\addplot graphics [xmin=947.368,xmax=3052.63,ymin=64.976,ymax=185.024] {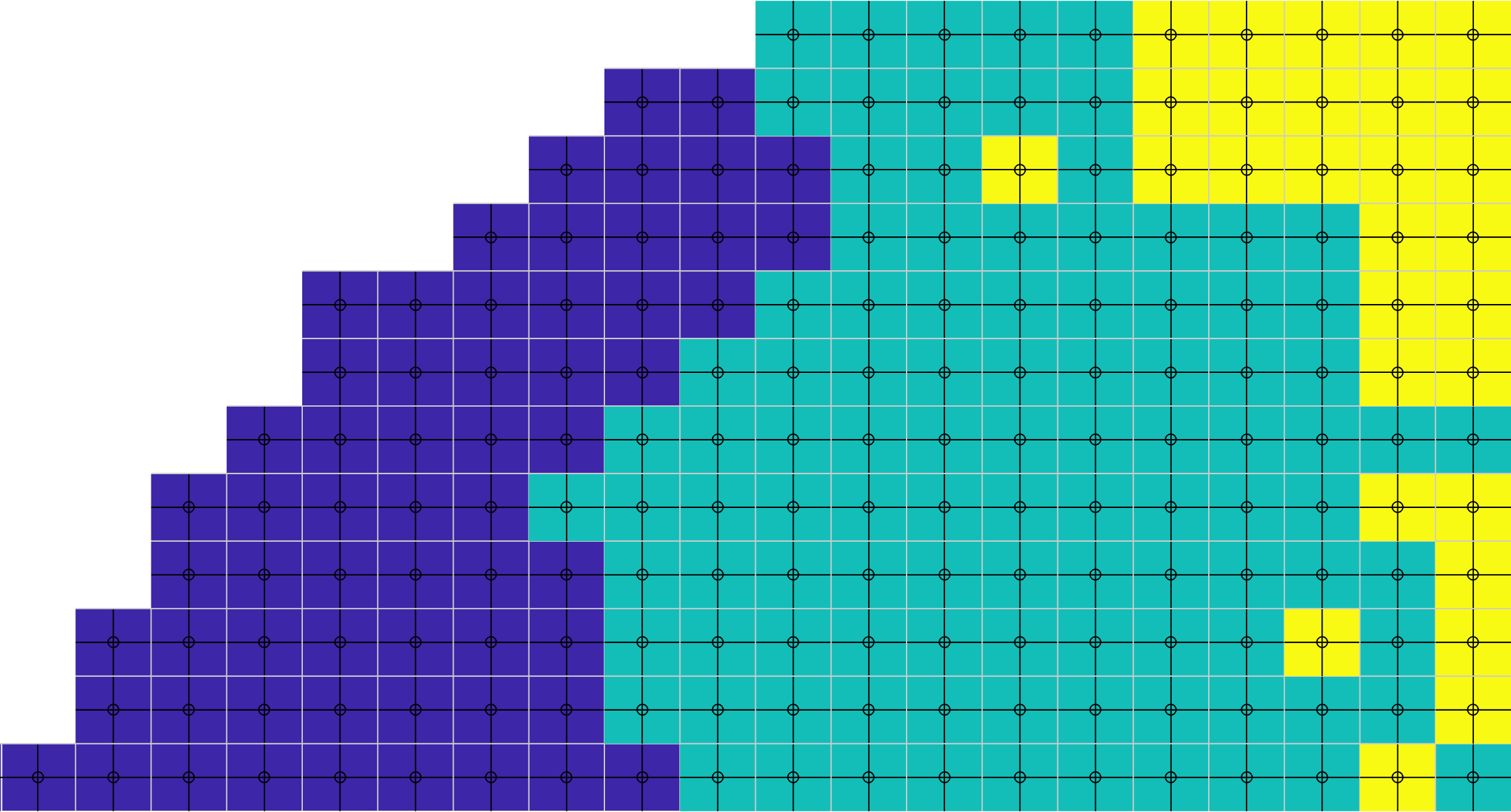};
\end{axis}
\node [left=of resFmfTwoAbsGammaMuPic.south west,  anchor=south east, shift=({0.2, -0.4})] {(a)};
\node [left=of resNrThreeAbsGammaMuPic.south west, anchor=south east, shift=({0.2, -0.4})] {(b)};
\end{tikzpicture}
    \caption{
        \cGls{fwm} bandwidths for \cglspl{fmf} and \cgls{nr} waveguides with enforced high nonlinearity coefficients.
        Bandwidths are normalized to multiples of the C-band's frequency range.
        Waveguides in white areas don't support enough guided modes.\\
        (a): \cGls{2fwm} in \cglspl{fmf}         under the constraint $\gamma >= \qty{1e-4}{\per\watt\per\meter}$.\\
        (b): \cGls{3fwm} in \cgls{nr} waveguides under the constraint $\gamma >= \qty{30}{\per\watt\per\meter}$.\\
        The figures for \cGls{1fwm} in \cglspl{fmf}, \cGls{1fwm} in \cgls{nr} waveguides, and \cGls{2fwm} in \cgls{nr} waveguides are exactly the same as \crefSubfig{fig:res_regular_mu}{a}, \crefSubfig{fig:res_regular_mu}{d} and \crefSubfig{fig:res_regular_mu}{e}, respectively, and are not repeated here.
        The bandwidths for \cGls{3fwm} in \cglspl{fmf} are always zero ($\gamma$ too small) and hence we don't include a figure.
    }
    \label{fig:res_absGamma_mu}
\end{figure}
\begin{figure}[tb]
    \centering
    \footnotesize
    \setlength\figW{0.8\linewidth} \begin{tikzpicture}
\begin{axis}[
    name=resFmfOneAbsGammaGammaPic,
    enlargelimits=false,
    axis on top,
    width=0.72\figW,
    height=0.35\figW,
    scale only axis,
    xticklabels={},
    ylabel={$w_\text{Trench} \, \mathrm{\left[\mu{}m\right]}$},
    ylabel style={font=\color{white!15!black}, yshift=-1mm},
    ytick={0, 1, 2, 4, 8},
    yticklabel style={text width=2em, align=right}, 
    point meta min=0.140248, point meta max=1.84082, 
    colorbar,
    colorbar style={
        ylabel={$\gamma$ [\unit{\per\watt\per\km}]},
        at={(1.03,0.5)},
        anchor=west,
        yticklabel style={/pgf/number format/fixed,},
        scaled y ticks=false,
        ytick style={color=black},
        tick pos=right,
        width=8,
    },
]
\addplot graphics [xmin=4.9928,xmax=41.0072,ymin=-0.12665,ymax=8.12665] {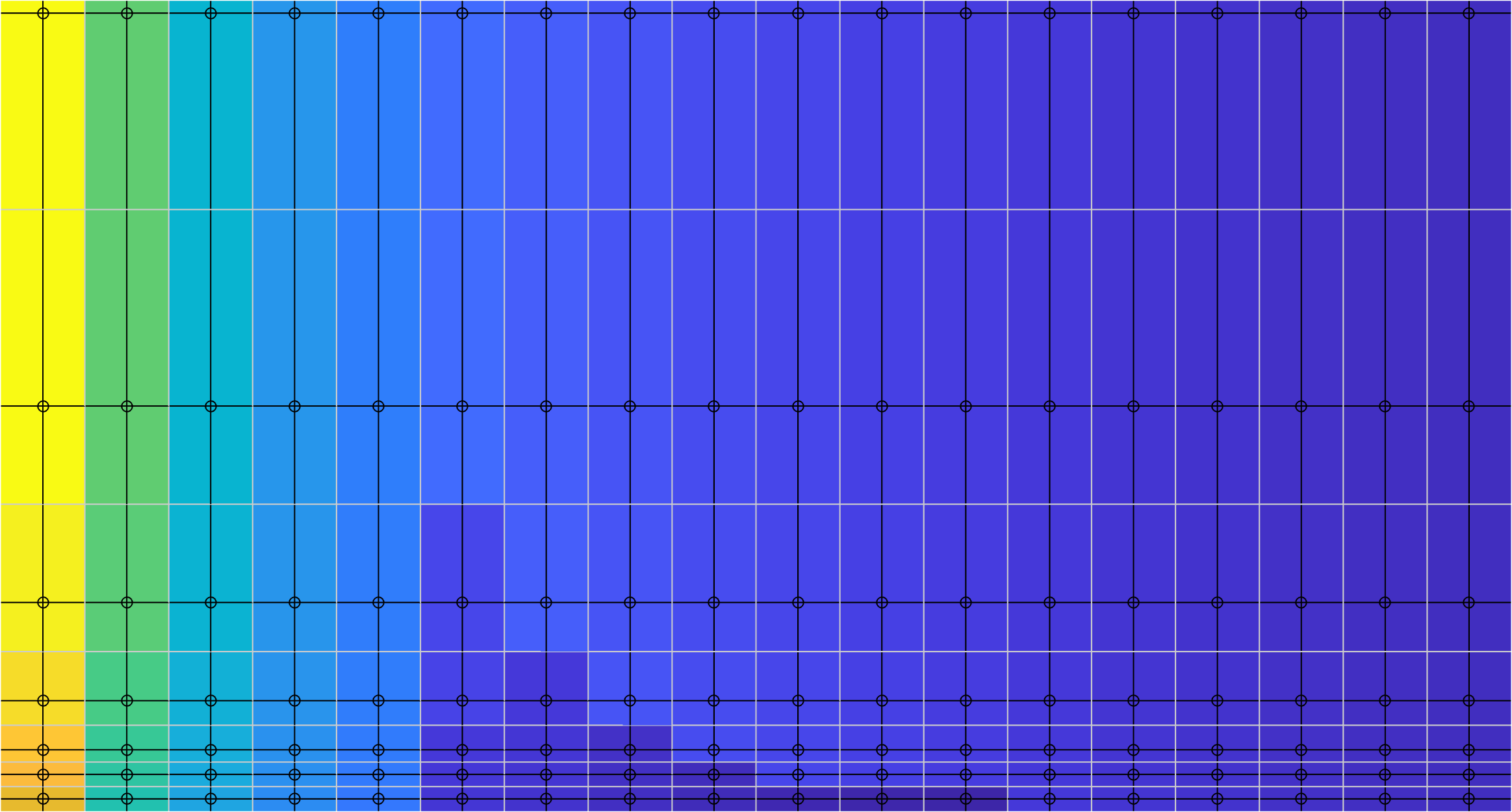};
\end{axis}
\begin{axis}[
    name=resFmfTwoAbsGammaGammaPic,
    at = (resFmfOneAbsGammaGammaPic.south),
    anchor = north,
    yshift = -0.2cm,
    enlargelimits=false,
    axis on top,
    width=0.72\figW,
    height=0.35\figW,
    scale only axis,
    xlabel style={font=\color{white!15!black}},
    xlabel={$r_\text{Core} \, \mathrm{\left[\mu{}m\right]}$},
    ylabel={$w_\text{Trench} \, \mathrm{\left[\mu{}m\right]}$},
    ylabel style={font=\color{white!15!black}, yshift=-1mm},
    ytick={0, 1, 2, 4, 8},
    yticklabel style={text width=2em, align=right}, 
    point meta min=0, point meta max=0.297809, 
    colorbar,
    colorbar style={
        ylabel={$\gamma$ [\unit{\per\watt\per\km}]},
        at={(1.03,0.5)},
        anchor=west,
        yticklabel style={/pgf/number format/fixed,},
        scaled y ticks=false,
        ytick style={color=black},
        tick pos=right,
        width=8,
    },
]
\addplot graphics [xmin=4.9928,xmax=41.0072,ymin=-0.12665,ymax=8.12665] {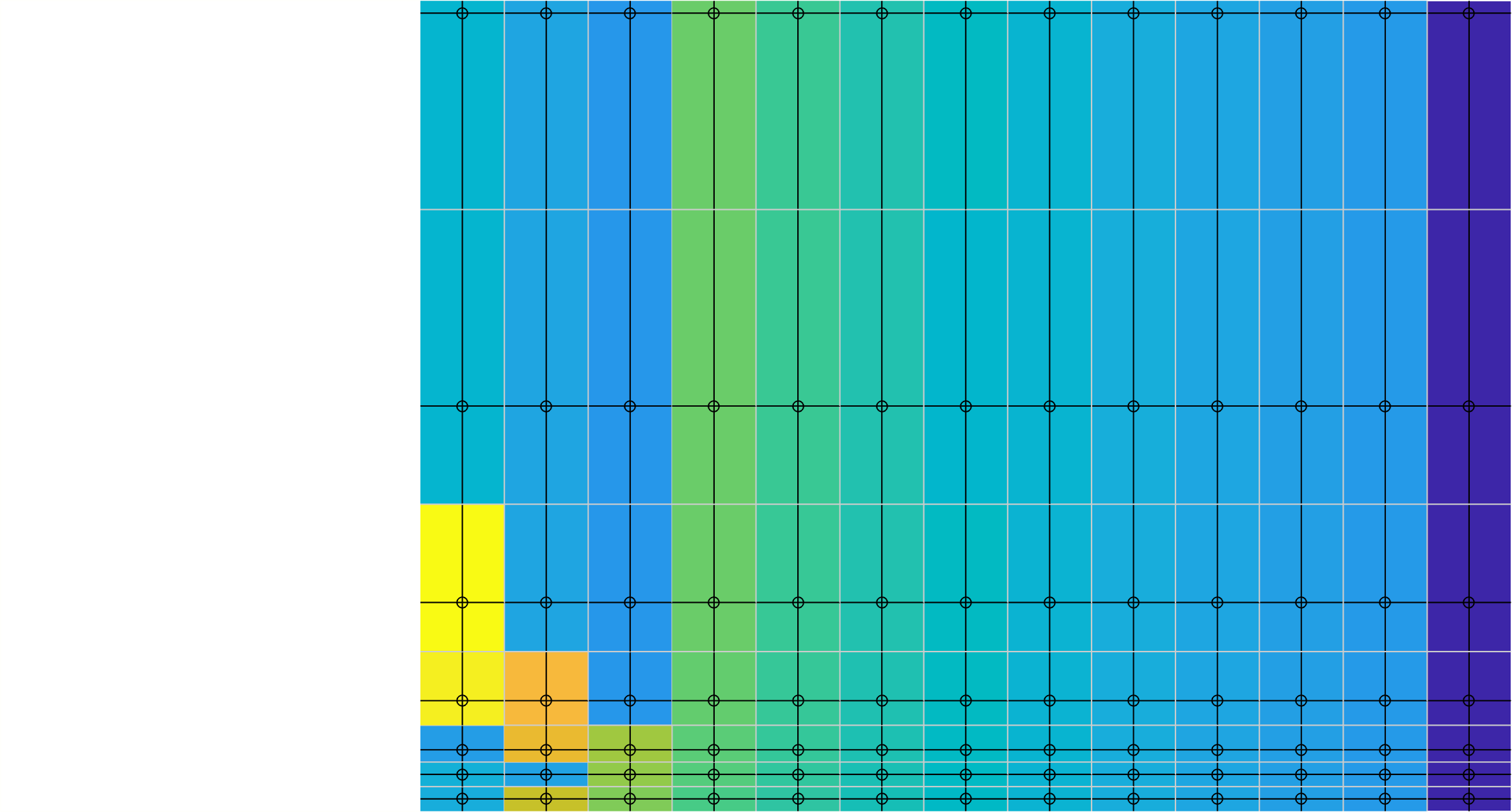};
\end{axis}
\begin{axis}[
    name=resNrOneAbsGammaGammaPic,
    at = (resFmfTwoAbsGammaGammaPic.south),
    anchor = north,
    yshift = -1cm,
    enlargelimits=false,
    axis on top,
    width=0.72\figW,
    height=0.35\figW,
    scale only axis,
    xticklabels={},
    ylabel={$h_\text{Slab} \, \mathrm{\left[nm\right]}$},
    ylabel style={font=\color{white!15!black}, yshift=-1mm},
    ytick={70, 90, 110, 130, 150, 170},
    yticklabel style={text width=2em, align=right}, 
    point meta min=95.5224, point meta max=305.105,
    colorbar,
    colorbar style={
        ylabel={$\gamma$ [\unit{\per\watt\per\meter}]},
        at={(1.03,0.5)},
        anchor=west,
        ytick style={color=black},
        tick pos=right,
        width=8,
    },
]
\addplot graphics [xmin=947.368,xmax=3052.63,ymin=64.976,ymax=185.024] {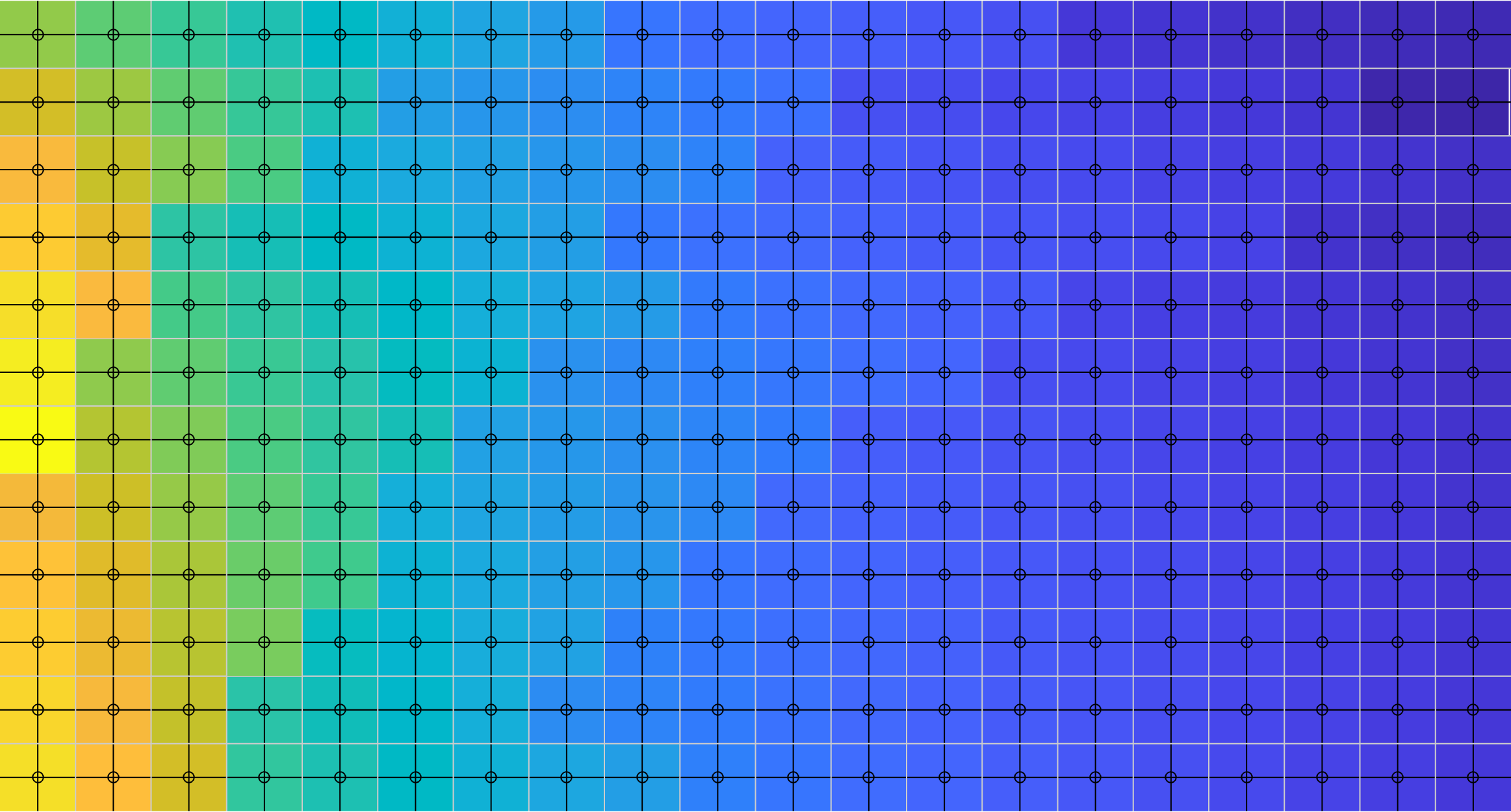};
\end{axis}
\begin{axis}[
    name=resNrTwoAbsGammaGammaPic,
    at = (resNrOneAbsGammaGammaPic.south),
    anchor = north,
    yshift = -0.2cm,
    enlargelimits=false,
    axis on top,
    width=0.72\figW,
    height=0.35\figW,
    scale only axis,
    xticklabels={},
    ylabel={$h_\text{Slab} \, \mathrm{\left[nm\right]}$},
    ylabel style={font=\color{white!15!black}, yshift=-1mm},
    ytick={70, 90, 110, 130, 150, 170},
    yticklabel style={text width=2em, align=right}, 
    point meta min=0, point meta max=180.589,
    colorbar,
    colorbar style={
        ylabel={$\gamma$ [\unit{\per\watt\per\meter}]},
        at={(1.03,0.5)},
        anchor=west,
        ytick style={color=black},
        tick pos=right,
        width=8,
    },
]
\addplot graphics [xmin=947.368,xmax=3052.63,ymin=64.976,ymax=185.024] {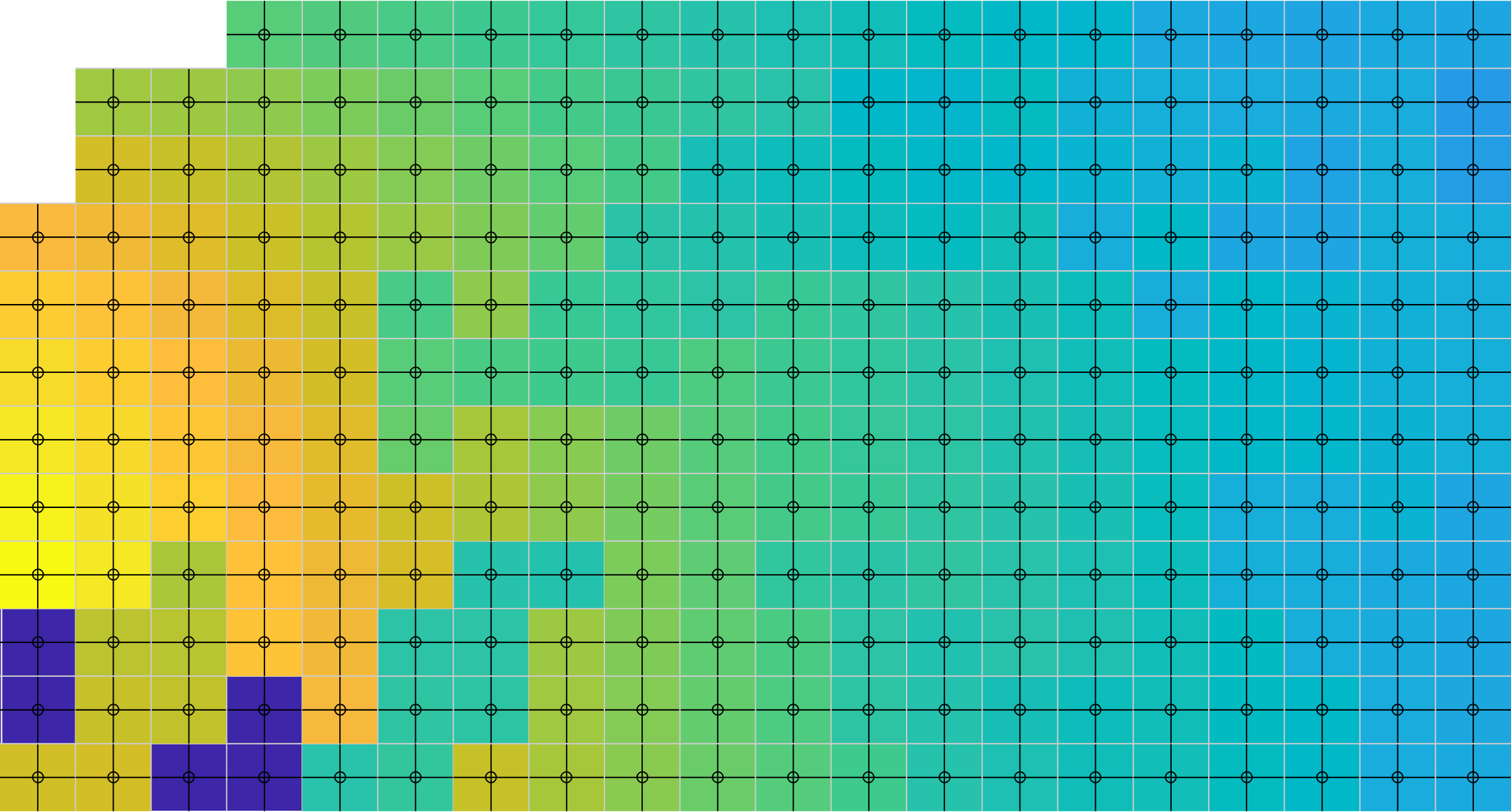};
\end{axis}
\begin{axis}[
    name=resNrThreeAbsGammaGammaPic,
    at = (resNrTwoAbsGammaGammaPic.south),
    anchor = north,
    yshift = -0.2cm,
    enlargelimits=false,
    axis on top,
    width=0.72\figW,
    height=0.35\figW,
    scale only axis,
    xlabel style={font=\color{white!15!black}},
    xlabel={$w_\text{Rib} \, \mathrm{\left[nm\right]}$},
    xtick={1000, 1421, 1842, 2157, 2578, 3000},
    ylabel={$h_\text{Slab} \, \mathrm{\left[nm\right]}$},
    ylabel style={font=\color{white!15!black}, yshift=-1mm},
    ytick={70, 90, 110, 130, 150, 170},
    yticklabel style={text width=2em, align=right}, 
    point meta min=0, point meta max=41.668,
    colorbar,
    colorbar style={
        ylabel={$\gamma$ [\unit{\per\watt\per\meter}]},
        at={(1.03,0.5)},
        anchor=west,
        ytick style={color=black},
        tick pos=right,
        width=8,
    },
]
\addplot graphics [xmin=947.368,xmax=3052.63,ymin=64.976,ymax=185.024] {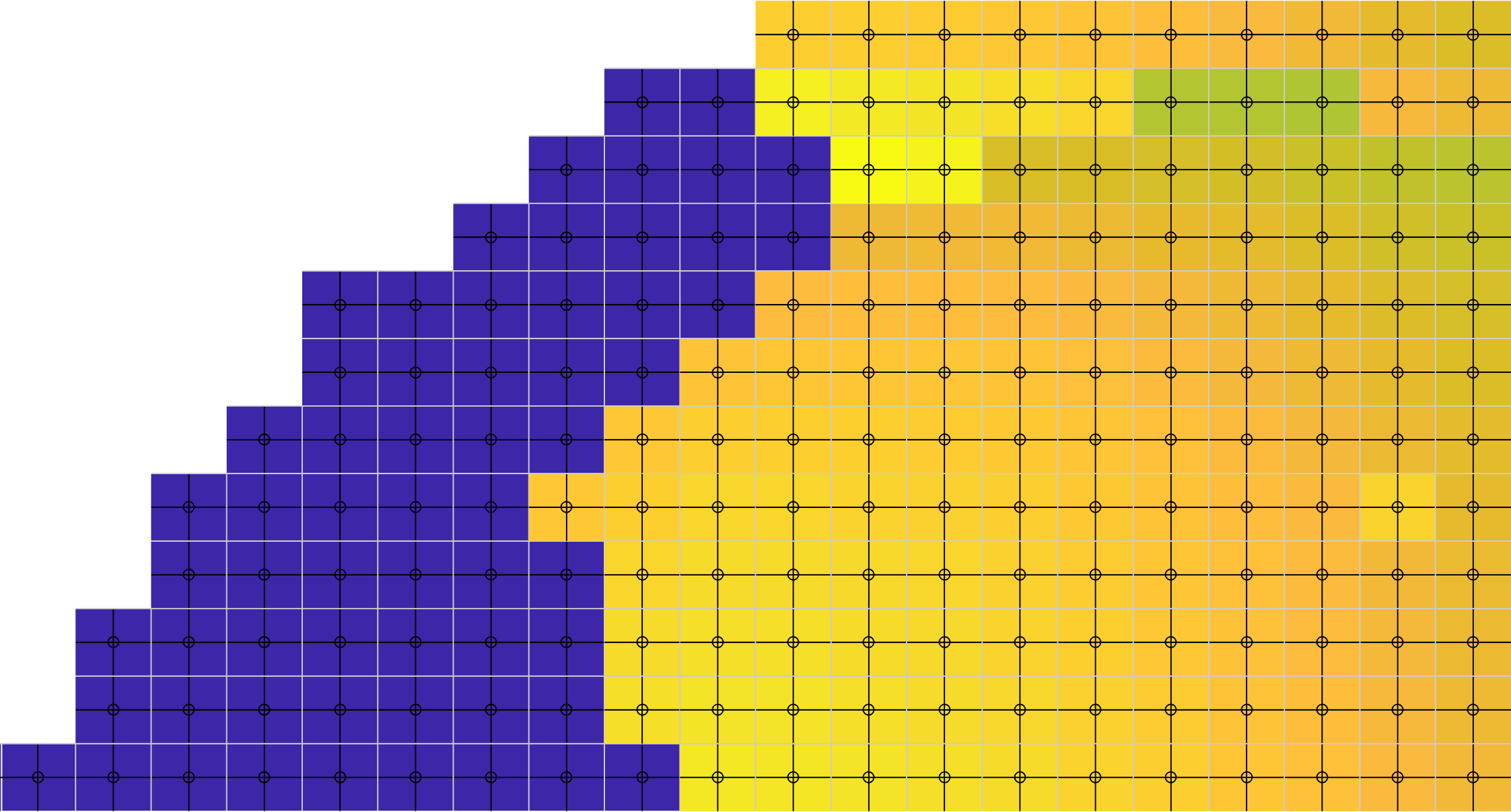};
\end{axis}
\node (a label) [left=of resFmfOneAbsGammaGammaPic.south west,  anchor=south east, shift=({0.2, -0.4})] {(a)};
\node (b label) [left=of resFmfTwoAbsGammaGammaPic.south west,  anchor=south east, shift=({0.2, -0.4})] {(b)};
\node (c label) [left=of resNrOneAbsGammaGammaPic.south west,   anchor=south east, shift=({0.2, -0.4})] {(c)};
\node (d label) [left=of resNrTwoAbsGammaGammaPic.south west,   anchor=south east, shift=({0.2, -0.4})] {(d)};
\node (e label) [left=of resNrThreeAbsGammaGammaPic.south west, anchor=south east, shift=({0.2, -0.4})] {(e)};
\end{tikzpicture}
    \caption{
        Approximate nonlinearity coefficients for the \rev{same configurations as in \cref{fig:res_absGamma_mu} (\cref{fig:res_regular_mu} for subfigures which are not repeated in \cref{fig:res_absGamma_mu}).}
        Waveguides in white areas don't support enough guided modes.\\
        (a): \cGls{1fwm} in \cglspl{fmf}.
        (b): \cGls{2fwm} in \cglspl{fmf}.
        (c): \cGls{1fwm} in \cgls{nr} waveguides.
        (d): \cGls{2fwm} in \cgls{nr} waveguides.
        (e): \cGls{3fwm} in \cgls{nr} waveguides.
        We don't include a figure for \cgls{3fwm} in \cglspl{fmf}, since the threshold $\gamma >= \qty{1e-4}{\per\watt\per\meter}$ is never fulfilled.
    }
    \label{fig:res_absGamma_gamma}
\end{figure}
\begin{figure}[tb]
    \centering
    \footnotesize
    \setlength\figW{0.7\linewidth} \input{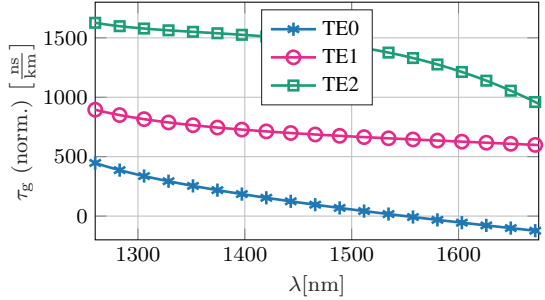}
    \caption{
        Relative normalized group delay of modes in a \cgls{nr} waveguide with
        $w_\text{Rib} = \qty{1210.53}{\nm}$ and $h_\text{Slab} = \qty{70}{\nm}$.
    }
    \label{fig:disp_nr_pmFail}
\end{figure}
From the results in \cref{sec:34PM}, one could conclude that \cgls{3fwm} gives almost equal results as \cgls{1fwm} or \cgls{2fwm} in the considered parameter ranges.
However, the power of a generated \cgls{fwm} idler also depends on the nonlinearity coefficient.
For this feasibility study, \rev{we} approximate the idler power as \cite{rademacherInvestigationIntermodalFourwave2018}
\begin{align}
    \mathrm{P_I} = 4 (\gamma^{\mA,\mB,\mC,\mD})^2 \LL^2 \PP_\mathrm{P_1} \PP_\mathrm{S} \PP_\mathrm{P_2} \rev{\expp{-\alpha \LL}} |\qfwmEff(\rev{\Delta\beta})|^2,\label{eq:idlerPower}
\end{align}
with waveguide length $\LL$, attenuation $\alpha$ and laser powers $\PP_\mathrm{P_1}$, $\PP_\mathrm{S}$, $\PP_\mathrm{P_2}$.
The nonlinearity coefficient is defined as \cite{rademacherInvestigationIntermodalFourwave2018,rademacherNonlinearGaussianNoise2016}
\begin{align}
\gamma^{\mA,\mB,\mC,\mD} = \frac{  \omega_0 \nn_2  }{  \cc_0 \mathrm{A_{eff,cross}^{\mA,\mB,\mC,\mD}}}\label{eq:gamma}
\end{align}
with nonlinear refractive index $\nn_2$ and speed of light in vacuum $\cc_0$.
The cross effective area between involved modes is given by \cite{rademacherNonlinearGaussianNoise2016}
\begin{align}
    \mathrm{A_{eff,cross}^{\mA,\mB,\mC,\mD}} = \frac{
        \sqrt{\mathcal{I}^\mA \mathcal{I}^\mB \mathcal{I}^\mC \mathcal{I}^\mD}
    }{
        \left| \iint    \left[ \left(\Vec{F}^{\mA}\right)^\conj \cdot \Vec{F}^{\mB} \right]    \left[ \left(\Vec{F}^{\mC}\right)^\conj \cdot \Vec{F}^{\mD} \right]     \mathrm{dA} \right|
    },\label{eq:AEffCross}
\end{align}
where \rev{$\Vec{F}^\mX$} are the transversal mode profiles, $(.)^\conj$ denotes complex conjugation, \rev{$\mathcal{I}^\mX = \iint \abs{\Vec{F}^\mX}^2 \mathrm{dA}$} and integrals are taken over the waveguide cross section.

We can fix $\omega_0$ in \cref{eq:gamma} to the same value for all optimizations, since it scales all results equally.
We selected the highest simulated frequency, i.e., the lowest wavelength of the O-band.
Since we are only interested in the approximate difference in idler powers between different geometries, we can safely neglect that the mode profiles in the integrals in \cref{eq:AEffCross} change with pump and signal frequencies.
Furthermore, for these integrals, we only consider regions \textit{inside} of the waveguide core.
This way, we avoid integrating over material borders where the electrical field shows Dirac-delta-like behavior.
Integrating over borders needs a much higher resolution and even then the results are distorted.
Finally, note that although the idler power depends on waveguide length, the nonlinearity parameter does not.
\rev{Despite having performed \cgls{fmf} \cgls{pm} with the artificially short $\LL = \qty{10}{\meter}$, we use a realistic fiber length of \qty{4.7}{\km} for computing the idler power in \cref{eq:idlerPower}.}

\Cref{fig:ol} shows two configurations of overlapped and multiplied fields (i.e., the integrand in the denominator of $\mathrm{A_{eff,cross}^{\mA,\mB,\mC,\mD}}$ in \cref{eq:AEffCross}) of an exemplary \cgls{nr} waveguide.
The black lines show field magnitudes over the horizontal dimension.
It can be seen that the integral over \crefSubfig{fig:ol}{a} will give a positive number and the integral over \crefSubfig{fig:ol}{b} will be close to zero, since positive and negative lobes cancel each other.
Therefore, the idler power will be much higher for the first case.
\rev{Computed nonlinearity coefficients in the example are \qty{20.2}{\per\watt\per\m} and \qty{0.0058}{\per\watt\per\m} -- a factor of more than \qty{35}{\dB}}.

In the following, we present \cgls{fwm} bandwidths and nonlinearity coefficients, but force the optimizer to only accept \cgls{fwm} configurations where nonlinearity coefficients have at least a predefined minimal value.
For \cglspl{fmf}, this value is $\gamma >= \qty{1e-4}{\per\watt\per\meter}$ and for \cgls{nr} waveguides $\gamma >= \qty{30}{\per\watt\per\meter}$.
We selected these numbers, since they represent roughly $10\%$ of the highest achievable values in their respective waveguide types.

\paragraph*{\textbf{Few-Mode Fibers}}
Firstly, enforcing $\gamma$ values above the chosen threshold does not change the results for \cgls{1fwm}, meaning that all configurations in \crefSubfig{fig:res_regular_mu}{a} have $\gamma$ values above the threshold.
Secondly, comparing \crefSubfig{fig:res_absGamma_mu}{a} with \crefSubfig{fig:res_regular_mu}{b} shows that enforcing a minimal $\gamma$ also does not have a huge effect on \cgls{2fwm}.
We see similar bandwidths, except for fibers with $r_\text{core} = \qty{40}{\um}$.
\rev{Note, however, that the configurations (wavelengths and mode choices) which lead to the values in \cref{fig:res_absGamma_mu} can differ from those leading to \cref{fig:res_regular_mu}, since they are the result of a completely new optimization.}
A bandwidth of zero in the figure means that our limit on $\gamma$ could not be achieved by any \cgls{fwm} configuration.
The reason is that larger cores have larger effective areas, which leads to lower nonlinearity (see \cref{eq:gamma}).
\rev{For \cgls{opc}, the maximal bandwidth is still \num{1.8}, as in \cref{fig:res_regular_mu}.}
Finally, there is no \cgls{3fwm} configuration in the \cglspl{fmf} considered here with $\gamma$ values above our limit.
Therefore, idler powers are very weak and we consider \cgls{3fwm} in \cglspl{fmf} as infeasible.

\CrefSubfig{fig:res_absGamma_gamma}{a} shows the resulting nonlinearity coefficients $\gamma$ for the constrained optimization \rev{(\cref{fig:res_absGamma_mu})}.
It is clear that larger cores have lower nonlinearity and that almost all waveguides have $\gamma$ values at least 5 times higher than our limit.
\CrefSubfig{fig:res_absGamma_gamma}{a} and \crefSubfig{fig:res_regular_mu}{a} reveal that waveguides with small cores have high \cgls{1fwm} bandwidth and high nonlinearity coefficients.
For \cgls{2fwm}, \crefSubfig{fig:res_absGamma_gamma}{b} shows that $\gamma$ values are lower (roughly around our limit) and the good \cgls{pm} configurations for large cores in \crefSubfig{fig:res_regular_mu}{b} have low nonlinearity.
However, the configurations with core sizes around $\qty{16}{\um}$ have good bandwidth, acceptable nonlinearity coefficients and \cgls{2fwm} is feasible in those fibers.

\paragraph*{\textbf{Nano-Rib Waveguides}}
Enforcing $\gamma$ values above the chosen threshold does not change the bandwidths in \cgls{1fwm} and \cgls{2fwm}.
Comparing \crefSubfig{fig:res_absGamma_mu}{b} with \crefSubfig{fig:res_regular_mu}{f} shows that \cgls{3fwm} is severely affected by enforcing high $\gamma$ values.
But, in contrast to \cglspl{fmf}, \cgls{nr} waveguides \textit{do} support \cgls{3fwm}.
However, the bandwidth is reduced by orders of magnitude to roughly \qty{250}{\GHz}, which still can be enough for processing a few optical channels.
\rev{For \cgls{opc}, the maximum bandwidth is slightly higher at \num{0.16}.}

\Cref{fig:res_absGamma_gamma} (c) and (d) show that almost all \cgls{nr} geometries support \cgls{fwm} configurations with high $\gamma$ values.
For \cgls{1fwm}, all configurations have nonlinearity coefficients above \qty{100}{\per\watt\per\meter}.
For \cgls{2fwm}, the coefficients are close to \qty{100}{\per\watt\per\meter} \rev{(\num{220} in \cgls{opc})}, except for some scattered geometries where \cgls{pm} fails.
To understand the reason for this failure, \cref{fig:disp_nr_pmFail} shows dispersion curves of one of those waveguides.
The group delay curves do not have any vertical overlap, which means \cgls{2fwm} is impossible (see \cref{eq:pmApproxTwoMode}).

Finally, \crefSubfig{fig:res_absGamma_gamma}{e}, shows that \cgls{nr} \cgls{3fwm} $\gamma$ values are not very sensitive to geometry variations and the coefficients are around \qty{40}{\per\watt\per\meter} \rev{(\num{60} in \cgls{opc})}.

\section{\rev{Idler Power Evolution}}\label{sec:propa}
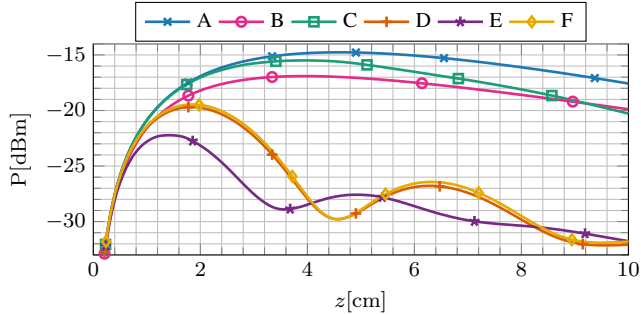
\begin{figure}[tb]
    \centering
    \footnotesize
    \resizebox{\linewidth}{!}{
        \setlength\figW{0.99\linewidth} \begin{tikzpicture}
\begin{axis}[%
    width = 0.99\figW,
    height = 0.5\figW,
    xmin = 0,
    xmax = 10,
    xlabel = {$z [\unit{\cm}]$},
    ymin = -33,
    ymax = -14,
    ylabel = {$\mathrm{P} [\unit{\dBm}]$},
    extra y ticks = {-14},
    extra y tick labels = {},
    minor tick num = 4,
    grid = both,
    mark repeat = {51},
    legend style = {
        legend columns = 6,
        at = {(0.5,1.01)},
        anchor = south,
    },
]
\addplot [color=colA, line width=1.0pt, mark size=2.0pt, mark=x, mark options={solid, colA}] table[]{\CurrentFilePath/src/propa_3000_140-1.tsv};
\addlegendentry{A}

\addplot [color=colB, line width=1.0pt, mark size=1.8pt, mark=o, mark options={solid, colB}] table[]{\CurrentFilePath/src/propa_3000_140-2.tsv};
\addlegendentry{B}

\addplot [color=colC, line width=1.0pt, mark size=1.7pt, mark=square, mark options={solid, colC}] table[]{\CurrentFilePath/src/propa_3000_140-3.tsv};
\addlegendentry{C}

\addplot [color=colD, line width=1.0pt, mark size=2.0pt, mark=+, mark options={solid, colD}] table[]{\CurrentFilePath/src/propa_3000_140-4.tsv};
\addlegendentry{D}

\addplot [color=colE, line width=1.0pt, mark size=2.0pt, mark=star, mark options={solid, colE}] table[]{\CurrentFilePath/src/propa_3000_140-5.tsv};
\addlegendentry{E}

\addplot [color=colF, line width=1.0pt, mark size=2.0pt, mark=diamond, mark options={solid, colF}] table[]{\CurrentFilePath/src/propa_3000_140-6.tsv};
\addlegendentry{F}
\end{axis}
\end{tikzpicture}%
    }
    \caption{
        \rev{Simulated evolution of the \cgls{bs} idler in the \cgls{nr} waveguide with rib width \qty{3000}{\nm} and slab height \qty{140}{\nm}, with five different propagation settings.
        See main text for settings values.}
    }
    \label{fig:propa_3000_140}
\end{figure}

\rev{The results in sections \cref{sec:34PM,sec:overlap} are based on the \cgls{fwm} efficiency $\qfwmEff$ \cref{eq:eta}.
This metric relies on three main approximations: 1) The attenuation is constant (neither frequency, nor mode dependent). 2) There is no linear coupling. 3) The pumps and signal are not depleted by the nonlinear interaction.
The third assumption is usually fulfilled in all-optical signal processing experiments, since the pumps and signal are much stronger (usually around \qty{20}{\dBm} and \qty{10}{\dBm}) than the generated idler (typically less than \qty{-20}{\dBm}).
However, due to the first two assumptions, $\qfwmEff$ fails to capture variable attenuation and linear coupling, which usually need to be considered in real-world applications.
The \cgls{fwm} efficiency can be remarkably exact, though, when linear coupling is low due to few propagating modes and the attenuation is flat due to small considered wavelength differences, e.g. in \cite{rademacherInvestigationIntermodalFourwave2018} or \cite{essiambreExperimentalInvestigationInterModal2013}.

In many of our presented configurations, attenuation is quite likely frequency dependent due to the large considered bandwidths.
Also, the waveguides we consider potentially allow the propagation of many guided modes, especially those with large cores.
Since the group delays of neighboring modes are very close in these waveguides (see e.g. \cref{fig:disp_fmf_large_core}), linear coupling will also have an effect on \cgls{fwm}.
Nevertheless, $\qfwmEff$ serves as a good metric for a best-case analysis -- given that modal overlap is considered as well (as in \cref{sec:overlap}).
For example, \cgls{fwm} efficiency predicted the bandwidth quite well in our \cgls{nr} waveguide-based all-optical C- to O-band wavelength conversion experiment \cite{ronnigerEfficientUltrabroadbandCtoO2021}.
However, linear crosstalk was much more severe in previous \cgls{nr} waveguide generations we had manufactured, prohibiting efficient conversion.

The interplay of variable attenuation, linear coupling and phase mismatch can lead to quite different nonlinear behavior (see e.g. \cite{xiaoTheoryIntermodalFourwave2014} or \cite[Ch.~3]{marhicFiberOpticalParametric2007}).
As an example, the idler evolution in an exemplary waveguide ($w_\text{Rib} = \qty{3000}{\nm}$ and $h_\text{Slab} = \qty{140}{\nm}$) is shown in \cref{fig:propa_3000_140} under different assumptions $\mathrm{A}$ to $\mathrm{F}$.
The power was computed with a frequency domain \cgls{cw} simulation including attenuation, linear coupling (considering modes \TE{0} to \TE{3}) and nonlinear coupling (based on \namecref{eq:eta} (1) in \cite{hoflerModelingMaterialSusceptibility2021}).
Here we used a constant nonlinear susceptibility $\chi^{[3]} = \complexqty{3.77 -j0,224e-19}{\V^2\per\m^2}$ or, equivalently, $n_2 = \qty{10e-18}{\m^2\per\W}$ and $\beta_\mathrm{TPA} = \qty{0.5e-11}{\m\per\W}$.
The figure shows \cgls{bs} idlers in a \cgls{3fwm} configuration where the signal was launched into mode \TE{2} at wavelength \qty{1502.0}{\nm} with power \qty{10}{\dBm}, pump 1 into \TE{2} at \qty{1316.0}{\nm} with \qty{20}{\dBm}, pump 2 into \TE{3} at \qty{1263.9}{\nm} with \qty{20}{\dBm} and the idler evolved in \TE{1} at \qty{1576.1}{\nm}.
These wavelengths and modes are the optimum which lead to the bandwidth shown in \cref{fig:res_absGamma_mu} for $w_\text{Rib} = \qty{3000}{\nm}$ and $h_\text{Slab} = \qty{140}{\nm}$.

In scenario $\mathrm{A}$, we assumed a flat attenuation of \qty{1}{\dB\per\cm}, no linear coupling and perfect \cgls{pm}.
In $\mathrm{B}$, we changed to mode- and frequency dependent loss -- ranging from \qty{0.56}{\dB\per\cm} to \qty{2.6}{\dB\per\cm}, increasing with frequency and mode order.
In $\mathrm{C}$, we added linear coupling which couples all modes.
Linear coupling is caused by random fluctuations of waveguide imperfections and we selected a model which leads to a pump crosstalk similar to \cite{ronnigerEfficientUltrabroadbandCtoO2021}.
Therefore, crosstalk in a real waveguide can be both, higher or lower than our selected value here.
In $\mathrm{D}$, we moved the signal frequency from it's optimum to the border of the \cgls{pm} region, which created a phase mismatch of $\Delta\beta = \qty{140}{\per\m}$.
The border was defined in \cref{thm:fwmBw} such that the idler should drop by \qty{3}{\dB} at the waveguide's end (\qty{2}{\cm}).
As one can see in the figure, the difference between $\mathrm{A}$ and $\mathrm{D}$ is indeed close to \qty{3}{\dBm} at \qty{2}{\cm}.
Scenario $\mathrm{E}$ combines variable attenuation, linear coupling and phase mismatch and one can see that it's behavior is different.
Finally, scenario $\mathrm{F}$ is like $\mathrm{D}$, but the two photon absorption coefficient was set to $\beta_\mathrm{TPA} = 0$ -- the effect is very small.

A waveguide length of \qty{2}{\cm} is not the optimum for this waveguide in terms of generated idler power.
However, our goal was to optimize and compare \cgls{fwm} bandwidths among different waveguide geometries and \cgls{fwm} types and hence we fixed the waveguide lengths -- allowing for a fair comparison.}

\section{Conclusions}
We showed that \cgls{1fwm} and \cgls{2fwm} perform better than \cgls{3fwm} in both, \cglspl{fmf} and \cgls{nr} waveguides.
\cGls{fwm} bandwidths as well as \rev{approximative} nonlinearity coefficients are both larger.
\rev{\cGls{3fwm} is achievable in some \cgls{nr} waveguides} with acceptable efficiency in narrowband operation.
Depending on the intended use case, it might be acceptable to sacrifice some idler power and broadband operation to gain the flexibility to place signal, pump and idler in three different modes.
\rev{\cGls{4fwm} does not work in neither \cglspl{fmf}, nor in \cgls{nr} waveguides.
Our results are a best-case analysis of \acrlong{pm} and actual \cGls{fwm} bandwidth can be smaller than the presented values.}



\bibliographystyle{IEEEtran}
\bibliography{zoteroBibtex}
\end{document}